\documentclass[aps,twocolumn,pra,floatfix]{revtex4-1}
\usepackage{placeins}
\usepackage{amsmath,amssymb,bm}
\usepackage{graphicx}
\usepackage{epstopdf}
\usepackage{latexsym}
\usepackage{subfigure}
\usepackage[usenames,dvipsnames]{color}
\usepackage{hyperref}
\usepackage{natbib}
\usepackage{color}
\usepackage{lipsum}

\hypersetup{    
  colorlinks,
  citecolor=Blue,
  linkcolor=Red,
  urlcolor=Blue}

\begin{document}
\newcommand{\cs}[1]{{\color{blue}$\clubsuit$#1}}

\newcommand{\ldc}[1]{{\color{blue}$\clubsuit$#1}}

\title{Internal Oscillations of a Dark-Bright Soliton in a Harmonic Potential}
\author{Majed O. D. Alotaibi and Lincoln D. Carr}
\affiliation{Department of Physics, Colorado School of Mines, Golden, CO 80401, USA}

\begin{abstract}
{We investigate the dynamics of a dark-bright soliton in a harmonic potential using a mean-field approach via coupled nonlinear Schr\"odinger equations appropriate to multicomponent Bose-Einstein condensates. We use a modified perturbed dynamical variational Lagrangian approximation, where the perturbation is due to the trap, taken as a Thomas-Fermi profile. The wavefunction ansatz is taken as the correct hyperbolic tangent and secant solutions in the scalar case for the dark and bright components of the soliton, respectively. We also solve the problem numerically with psuedo-spectral Runge-Kutta methods. We find, analytically and numerically, for weak trapping the internal modes are nearly independent of center of mass motion of the dark-bright soliton. In contrast, in tighter traps the internal modes couple strongly to the center of mass motion, showing that for dark-bright solitons in a harmonic potential the center of mass and relative degrees of freedom are not independent.  This result is robust against noise in the initial condition and should, therefore, be experimentally observable.}
\end{abstract}
\maketitle


\section{Introduction}
\label{sec:FRHPRA:Introduction}


Solitons are emergent excitations of atomic matter waves in Bose-Einstein condensates (BECs).  In their simplest form they appear in highly visible form as density peaks (bright soliton) or notches (dark solitons) in scalar BECs~\cite{dauxois2006physics,Kevrekidis2008,Kevrekidis2015,Pethick2008}. The experimental realization of multiple-component BECs, where different atom species or internal states of the same atom type can be populated, has aroused considerable interest in vector solitons. The two-component vector soliton of different forms (i.e., dark-dark solitons~\cite{Cuevas2012,Liu2012a, Hoefer2011b}, bright-bright solitons~\cite{Xun-Xu2011} or dark-bright solitons~\cite{Becker2008b,Busch2001,Frantzeskakis2012,Hamner2011b,Rajendran2009b,Zhang2009b}) give rise to much richer phenomena than the single-component BECs, where one already finds, for example, soliton trains~\cite{Strecker2002b}, domain walls~\cite{Filatrella2014}, collective excitations and complex dynamics. In this Article we focus on the case of the dark-bright soliton. Although in scalar BECs the bright soliton can only exist for attractive interatomic interactions~\cite{doi:10.1142/S0217979205032279}, it can also be induced in purely repulsive multi-component BECs when a second component is occupies the density notch formed by a dark soliton in the first component. In this way, a dark soliton in one component forms an effective potential that traps the bright soliton component and therefore allows the creation of a nonlinear excited state. These solitons are sometimes referred to as symbiotic.  We use the term \emph{dark-bright soliton} for clarity~\cite{Perez-Garcia2005,Frantzeskakis2012}.

The nonlinear Schr\"odinger equation (NLSE) without the potential term is an integrable equation and possess solitonic solutions. By adding a potential term, in our case a harmonic potential, we work with the celebrated Gross-Pitaevskii equation (GPE). The oscillation of nonlinear excitations in a harmonic potential is a common problem that has been the focus of many studies, as such large scale motions are easily observable in BEC experiments. Of particular interest is the oscillation of two-component excitation like a bright-bright soliton, dark-dark soliton~\cite{PhysRevA.97.043621} or dark-bright soliton~\cite{PhysRevA.84.053626}. In these studies, usually, the ansatz used to describe the dark-bright soliton contains one variable to describe the position of the dark and bright components. A more realistic situation is to relax this restriction and allow the two components to move freely by adding one more degree of freedom to the problem, namely, the internal oscillation between the two components. We study the coupling between the internal oscillation of the two components in the dark-bright soliton and the oscillation of the whole system in a harmonic potential. The harmonic potential modifies the background of the dark component in a dark-bright soliton. Therefore a Thomas-Fermi background approximation is needed where the new dark component wave function is represented by subtracting the old dark component density from the harmonic potential function.  The result is a dark soliton on a top of parabola-shaped background, Fig.~\ref{fig:FRHPRA:osc_DS_sketch}.

It is well-known in the classical two-body problem that relative and center of mass degrees of freedom are independent in a harmonic potential.  A dark-bright soliton represents an emergent two-body semiclassical object in the context of the mean-field approximation on the many-body wavefunction underlying the BEC.  To what extent does this emergent structure have the same properties as a classical two-body problem?  An elementary consideration is separation of relative and center of mass degrees of freedom.  Previous treatments have avoided this question by pinning the dark and bright solitons to the same position.  By relaxing this constraint, in this Article, via both variational Lagrangian analytical methods and numerical solution of the GPE, we show that in general relative and center-of-mass degrees of freedom are not independent for the dark-bright soliton.  In contrast, these degrees of freedom are independent in the uniform case, where the center-of-mass motion is associated with a Goldstone mode~\cite{majed2017}.  For a weak enough trap, the separation of variables from the uniform case is only very weakly affected by the trap.  However, as the trap strength grows this separation of variables is lost.

This Article is structured as follows. In Sec.~\ref{sec:FRHPRA:Analytical Calculations} we present the two-component GPE, the variational Lagrangian model, use perturbation theory, and derive the equations of motion for the bright and dark soliton components. In Sec.~\ref{sec:FRHPRA:Numerical calculations} we numerically integrate the dimensionless GPE using a psuedo-spectral Runge-Kutta method and study the dynamics of the oscillation of the dark-bright soliton in a harmonic potential. Finally, in Sec.~\ref{sec:FRHPRA:Conclusions} we summarize our conclusions.

\section{Analytical Calculations}
\label{sec:FRHPRA:Analytical Calculations}
\subsection{Lagrangian density and ansatz}
\label{sec:FRHPRA:Lagrangian density and ansatz}

The two-component dark-bright soliton is governed by coupled GPEs~\cite{Kevrekidis2008}, which describe the evolution of the macroscopic wave functions of Bose condensed atoms:
\begin{align}
\label{eq:1DSE}
i \hbar \frac{\partial}{\partial \tilde{t}} \tilde{u}  &= -\frac{\hbar^2}{2 m}\frac{\partial^2 \tilde{u}}{\partial \tilde{x}^2} +\biggl[  \tilde{g}_{1} |\tilde{u}|^2
 - \tilde{u}^2_{0} + \tilde{g} |\tilde{v}|^2 +\tilde{V}(\tilde{x}) \biggl] \tilde{u}, \nonumber \\
i \hbar \frac{\partial}{\partial \tilde{t}} \tilde{v} &= -\frac{\hbar^2}{2 m}\frac{\partial^2 \tilde{v}}{\partial \tilde{x}^2} +\biggl[  \tilde{g}_{2} |\tilde{v}|^2 +\tilde{g} |\tilde{u}|^2 +\tilde{V}(\tilde{x}) \biggl] \tilde{v},
\end{align}
%
where tildes denote dimensional quantities. The wave function of the dark soliton is given by $\tilde{u} \equiv \tilde{u}\left(\tilde{x},\tilde{t}\right)$  and of the bright soliton by $ \tilde{v}\equiv\tilde{v}\left(\tilde{x},\tilde{t}\right)$. The dark soliton wave function is rescaled to remove the background contribution, $\tilde{u_0}$~\cite{Kivshar1995}. Although this is not necessary for the harmonic trap since there is no divergence in the total number of atoms, in order to match smoothly onto the untrapped limit and connect well with previous results from a uniform system~\cite{majed2017}, we include this subtraction. The interaction strength, $\tilde{g}_{j} = 2  a_{j} N \hbar \omega_{\bot}$ for $\left(j=1,2\right)$, is renormalized to 1D~\cite{Carr2000a} where $\tilde{g}_{1}$ ($\tilde{g}_{2}$) represents the intra-atomic interaction for the dark (bright) component and $g$ is the inter-atomic interaction between the two components of the BEC. The total number of atoms is $N$, the scattering length is $a_{j}$ and $\omega_{\bot}$ is the oscillation frequency of the transverse trap. To nondimensionlize Eqs.~\eqref{eq:1DSE} we multiply them by $\left(\hbar \omega_{\bot}\right)^{-1}$ and scale all quantities according to the following units:
%
%
\begin{equation}
	\label{eq:scaling_units}
\begin{aligned}
x & = \frac{\tilde{x}}{\ell_{\bot}}, \\
t & =  \tilde{t} \omega_{\bot}, \\
g_{ij} & = \frac{\tilde{g}_{ij}}{\ell_{\bot} \hbar \omega_{\bot}},  \\
|u|^2 & = \ell_{\bot} |\tilde{u}|^2 ,\\
|v|^2 & = \ell_{\bot} |\tilde{v}|^2 ,\\
V(x) & = \frac{\tilde{V}(\tilde{x})}{\hbar \omega_{\bot} } ,\\
u^2_{0} & = \frac{\tilde{u}^2_{0}}{\hbar \omega_{\bot} } ,
\end{aligned}
\end{equation}
where $\ell_{\bot}=\sqrt{\hbar /\left(m \omega_{\bot}\right)}$ is the transverse harmonic oscillator length.

The dimensionless version of the coupled GPEs is,
\begin{align}
	\label{eq:coupled_NLSE_hamonic_potential}
i \frac{\partial}{\partial t} u & = -\frac{1}{2} \frac{\partial^2}{\partial x^2} u + V(x) u + \left[g_{1}|u|^2 +g |v|^2 - u^2_{0} \right] u, \nonumber \\
i \frac{\partial}{\partial t} v & = -\frac{1}{2} \frac{\partial^2}{\partial x^2} v + V(x) v + \left[g_{2}|v|^2 +g |u|^2  \right] v,
\end{align}
%

\begin{figure}
	\centering
	\includegraphics[width=\columnwidth]{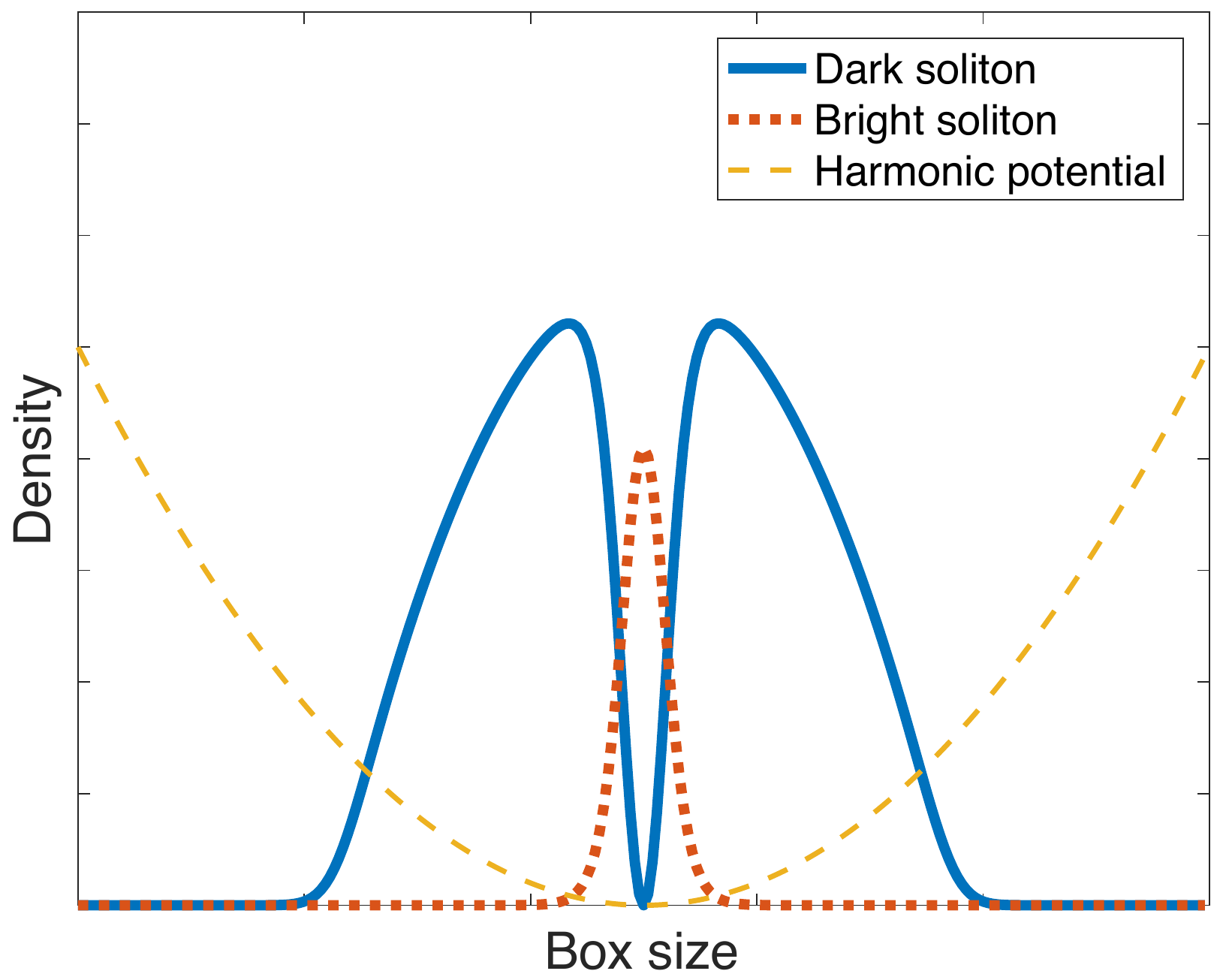}
	\caption{  \emph{Dark-bright soliton in harmonic potential well}. The background is affected by the harmonic trap, and therefore we work with the modified Thomas-Fermi cloud as described by Eq.~\eqref{eq:TF_harmonic}.}
	\label{fig:FRHPRA:osc_DS_sketch}
\end{figure}

The potential in Equations~\ref{eq:coupled_NLSE_hamonic_potential} takes the form,
\begin{align}
	\label{eq:potential1}
 V(x) = \frac{1}{2} \Omega^2 x^2, 
\end{align}
for both components. We assume $\Omega \ll 1$ and therefore we treat the harmonic potential as a small perturbation effect. Despite the fact that $x^2 \to \infty$ in Eq.~\eqref{eq:potential1}, because $V(x)$ always multiplies a background Thomas-Fermi wavefunction, the total perturbation is always small.  Even outside the Thomas-Fermi approximation, Gaussian tails in realistic BEC profiles in a harmonic trap will fall away much faster than $x^2$ diverges, making this perturbative picture a physically reasonable one beyond our approximations. The existence of the harmonic potential affects the background density of the dark-bright soliton, Fig.~\ref{fig:FRHPRA:osc_DS_sketch}. Consequently, we have to modify the usual assumption for a dark soliton of a uniform background and assume the dark soliton is supported by a Thomas-Fermi background condensate of form
\begin{align}
	\label{eq:TF_harmonic}
 |u_{\mathrm{TF}}|^2 = u^2_{0} - V(x).
\end{align}
We recast Eqs.~\eqref{eq:coupled_NLSE_hamonic_potential} to the following:
\begin{align}
	\label{eq:coupled_NLSE_hamonic_potential_rescaled}
& i \frac{\partial }{\partial t}u +\frac{1}{2} \frac{\partial^2 }{\partial x^2}u - \left[g_{1} \left| u \right|^2 +g \left| v \right|^2 -u^2_{0} \right] u = R_{u} \\ \nonumber
& i \frac{\partial }{\partial t}v +\frac{1}{2} \frac{\partial^2 }{\partial x^2}v - \left[g_{2} \left| v \right|^2 +g  \left| u \right|^2  \right] v = R_{v},
\end{align}
where the RHS of Eqs.~\eqref{eq:coupled_NLSE_hamonic_potential_rescaled} represent the perturbation effects,
\begin{align}
	\label{eq:RHS_perturb_harmonic}
& R_{u}= \frac{1}{2 u^2_{0}} \left[ 2 u (u^2_{0}-g_{1}|u|^2) V(x) + V^{\prime}(x) \partial_{x} u \right]   \\ \nonumber
& R_{v}= \frac{V(x)}{u^2_{0}}  \left[( u^2_{0} - g |u|^2) v \right].
\end{align}
Here $V^{\prime}(x) \equiv \frac{dV\left(x\right)}{dx} $. The Lagrangian density for the system of coupled equations, Eqs.~\eqref{eq:coupled_NLSE_hamonic_potential_rescaled} is:
%
%
%
\begin{align}
	\label{eq:LagrangiandDensity_harmonic}
\begin{split}
{\mathcal{L}} = & \frac{i}{2} \left[u^* \frac{\partial u}{\partial t} -u \frac{\partial u^*}{\partial t}  \right] \left[1-\frac{u^2_{0}}{g_{1}\left| u \right|^2} \right] - \frac{1}{2} \left|  \frac{\partial u }{\partial x}  \right|^2  \\
& - \frac{1}{2} \left[\sqrt{g_{1}} \left|u\right|^2 - \frac{u^2_{0}}{\sqrt{g_{1}}} \right]^2 + \frac{i}{2} \left[v^* \frac{\partial v}{\partial t} -v \frac{\partial v^*}{\partial t}  \right] \\
& - \frac{1}{2} \left|  \frac{\partial v }{\partial x}  \right|^2 -\frac{g_{2}}{2} \left| v \right|^4  - g \left| u \right|^2 \left| v \right|^2 .
\end{split}
\end{align}

We adopt the following trial functions as the dark-bright soliton solutions to Eqs.~\eqref{eq:coupled_NLSE_hamonic_potential_rescaled}:
\begin{align}
\label{eq:ansatz_harmonic}
u\left(x,t\right) &= \frac{u_{0}}{\sqrt{g_{1}}} \left\{ i A\left(t\right) + c\left(t\right)\ \mathrm{tanh}\left[\frac{\left(d\left(t\right) + x \right)}{\mathrm{w\left(t\right)}} \right] \right\}, \nonumber \\
v\left(x,t\right) &= \frac{u_{0}}{\sqrt{g_{2}}} F\left(t\right) \ \mathrm{sech} \left[\frac{\left(b\left(t\right) + x \right)}{\mathrm{w\left(t\right)}} \right] \\ \nonumber & \times   \mathrm{exp}{\left\{i \left[\phi_{0}\left(t\right) + x \phi_{1}\left(t\right) \right]\right\}}.
\end{align}
The parameters $A$, $c$, $F$ describe the amplitude of the two components where,
\begin{align}
	\label{eq:Ac}
 A^2+c^2=1,
\end{align}
and $A$ determines the velocity of the dark soliton component.  In the exponential term in Eqs.~\ref{eq:ansatz_harmonic}, $\phi_{0}$ gives rise to a complex amplitude to the bright soliton component. The velocity of the bright soliton is given by $\phi_{1}$, and $d$ and $b$ are the position of the dark and bright soliton, respectively. Since we are using hyperbolic functions as an ansatz, we assume the two components have the same width, $\mathrm{w}$,  for the problem to remain analytically tractable~\cite{Malomed1998}. There are 8 variational parameters subject to 1 constraint. The 8 variational parameters as shown in Eq.~\eqref{eq:ansatz_harmonic} are $A$, $c$, $d$, $\mathrm{w}$, $F$, $b$, $\phi_{0}$ and $\phi_{1}$ where we note Eq.~\eqref{eq:Ac} effectively reduces the number to 7.

In this ansatz, we have assumed a fixed background, i.e., there is no motion of the Thomas-Fermi background with respect to the harmonic trap.  The ansatz also neglects phonon effects.  Both of these restrictions will be relaxed in our numerical treatment in Sec.~\ref{sec:FRHPRA:Numerical calculations}.  We utilize the following normalization conditions,

\begin{subequations}
	\label{eq:normalizations}
	\begin{align}
	&\int_{-\infty}^{\infty} dx\; \left(\frac{u^2_{0}}{g_{1}} - \left|u\right|^2\right)=\frac{N_{1}}{N}, \\
	&\int_{-\infty}^{\infty} dx\; \left| v\right|^2 =\frac{N_{2}}{N}.
\end{align}
\end{subequations}
Here $N_1$ is the number of atoms displaced by the dark soliton and $N_2$ is the number of atoms in the bright soliton, and $N$ the total number of holes and atoms involved in the emergent feature of the dark-bright soliton only. In contrast, the total number of atoms in the condensate is $N_\mathrm{total}=\int dx |v|^2 + \int dx |u_{\mathrm{TF}}|^2 |u|^2 $. In general, $N_2 \ll N_\mathrm{total}$ as many more atoms are in the Thomas-Fermi background supporting the dark soliton, see the sketch in Fig.~\ref{fig:FRHPRA:osc_DS_sketch}.  Likewise the number of holes, i.e. the atoms displaced by the dark soliton, is typically much less than the total number of atoms even after subtracting out $N_2$, i.e., $N_1 \ll N_\mathrm{total}-N_2$.  This choice corresponds to the same normalization choice as used in unbounded systems without traps, and therefore allows us to check all results in the limit that trap frequency $\Omega \to 0$. By inserting the ansatz, Eqs.~\eqref{eq:ansatz_harmonic}, in the normalization, Eqs.~\eqref{eq:normalizations}, we find the relation between $N_{1}, N_{2}$ and the coefficients of the two components in the dark-bright soliton:
\begin{subequations}
	\label{eq:relation1}
\begin{align}
\frac{2 u^2_{0} c^2 w}{g_{1}} = \frac{N_{1}}{N}, \\
\frac{2 u^2_{0} F^2 w}{g_{2}} = \frac{N_{2}}{N}, \\
N=N_{1}+N_{2}
\end{align}
\end{subequations}

Out of the 8 experimental parameters $g, N, g_{1}, N_{1}, g_{2}, N_{2}, u_{0} $ and $ \Omega$, only 5 remain after taking into account the 3 constraints of Eqs.~\eqref{eq:relation1}  after the variational procedure. We choose $g, N_1/N_2, g_{1}, g_{2}$ and $ \Omega$ as the ``free parameters".

\subsection{Evolution equations}
\label{sec:FRHPRA:Evolution equations}

Using a perturbation technique in the variational method also modifies the standard Euler-Lagrange. To find the equations of motion that govern the behavior of the variational parameters we utilize the following modified Euler-Lagrange equation as defined in~\cite{Kivshar1995}:
\begin{equation}
	\label{eq:modified_Euler_Lagrange_eq}
	\frac{\partial L}{\partial a_{j}}  - \frac{d}{dt} \left(\frac{\partial L}{\partial \dot{a}_{j}}  \right) = 2\ \text{Re} \{ \int_{-\infty}^{\infty}( R^*_{u} \frac{\partial u }{\partial a_{j}} + R^*_{v} \frac{\partial v }{\partial a_{j}} ) \;dx  \}.
\end{equation}
Here $L = \int_{-\infty}^{\infty} dx \mathcal{L}$, $\mathcal{L}$ is the Lagrangian density in Eq.~\eqref{eq:LagrangiandDensity_harmonic} and $a_{j}$ represents the variational parameters where $\dot{a}_{j} \equiv da/dt $. We obtain $R^*_{u}$ and $R^*_{v}$ by inserting Eq.~\eqref{eq:ansatz_harmonic} into Eq.~\eqref{eq:RHS_perturb_harmonic} and take the conjugate of the outcome. Also, inserting Eqs.~\eqref{eq:ansatz_harmonic} into Eq.~\eqref{eq:LagrangiandDensity_harmonic} and integrating, we obtain the Lagrangian as a function of the variational parameters,
\begin{align}
	\label{eq:Lagrangian_fun}
L &= -\frac{2 u^2_{0} c^2}{3 g_{1} \mathrm{w}}-\frac{u^2_{0} F^2}{3 g_{2} \mathrm{w}}-\frac{2 u^4_{0} c^4 \mathrm{w}}{3 g_{1}}
+ \frac{2 g u^4_{0} (-1+c^2) F^2  \mathrm{w}}{g_{1} g_{2}} \nonumber \\ \nonumber &
-\frac{2 u^4_{0} F^4 \mathrm{w}}{3 g_{2}}
+ \frac{g u^4_{0} c^2 F^2}{g_{1} g_{2}} \mathrm{csch}\left(\frac{b - d }{\mathrm{w}}\right)^2 \\ \nonumber & \times
 \left\{ 4 \; \mathrm{coth}\left(\frac{b - d }{\mathrm{w}}\right) \left(b-d\right) - \left[3+ \mathrm{cosh}\left(2\frac{b - d }{\mathrm{w}}\right) \mathrm{w} \right] \right\}
\\ \nonumber & -\frac{u^2_{0} F^2 \mathrm{w} \phi^2_{1}}{g_{2}} - \frac{2 u^2_{0} }{g_{1}}  \left[\mathrm{tan}^{-1}\left(\frac{c}{A}\right)-A c\right] \frac{d}{dt} d
\\& -\frac{2u^2_{0} F^2 \mathrm{w} }{g_{2}} \frac{d}{dt} \phi_{0} +\frac{2u^2_{0} b  F^2 \mathrm{w} }{g_{2}} \frac{d}{dt} \phi_{1}.
\end{align}

Applying the modified Euler-Lagrange equations, Eq.~\eqref{eq:modified_Euler_Lagrange_eq}, yields a system of coupled nonlinear ordinary differential equations (ODEs) that describe the evolution in time of the variational parameters under the influence of the harmonic potential,
%
%
\begin{subequations}
		\label{eq:evolution_eqs}
\begin{align}
	\label{eq:evolution_eqs_1}
&\dot{\phi_{1}} = \frac{g u^2_{0} c^2}{g_{1} \mathrm{w}} \Gamma_{1} +\frac{\Omega^2}{45 g_{1}} \left\{b \left[45(g_{1}-g) \right. \right. \\ \nonumber & \left. \left.  -g(\pi^2-15)c^2+g \pi^2 c^2 d\right]\right\}	\\
 \label{eq:evolution_eqs_2}
&\dot{A} =  \frac{g u^2_{0} c F^2}{2 g_{2} \mathrm{w}} \Gamma_{1} + \frac{c(2+u^2_{0} c^2 \mathrm{w}^2)d}{6 u^2_{0} \mathrm{w}}  \Omega^2 \\
\label{eq:evolution_eqs_3}
&\frac{2 u^2_{0} c}{g_{1}}\dot{d} + \frac{2 u^2_{0} c}{g_{1} A^2 \left(1+\frac{c^2}{A^2}\right)}\dot{d} =
\frac{\pi^2 u^2_{0} A c^2 \mathrm{w}^3}{6 g_{1}} \Omega^2 \\
\label{eq:evolution_eqs_4}
&\frac{2 u^2_{0} A}{g_{1}}\dot{d} - \frac{2 u^2_{0}}{g_{1} A \left(1+\frac{c^2}{A^2}\right)}\dot{d} = \\ \nonumber &
\frac{c \mathrm{w} \left[18+(12+\pi^2) u^2_{0} c^2 \mathrm{w}^2 \right]}{18 g_{1}} \Omega^2 - \frac{2 g u^4_{0} c F^2 \Gamma_{2}}{g_{1}g_{2}}
\\ \nonumber &
-\frac{4 u^2_{0} c}{3 g_{1} g_{2} \mathrm{w}} \left[-g_{2}+u^2_{0}\left(-2 g_{2} c^2+3 g F^2\right) \mathrm{w}^2\right] \\
\label{eq:evolution_eqs_5}
& 2 \mathrm{w} \dot{F} + F \dot{\mathrm{w}} =0 \\
\label{eq:evolution_eqs_6}
& F \mathrm{w} \left(\phi_{1}+\dot{b}  \right) + b \left(2 \mathrm{w} \dot{F} + F \dot{\mathrm{w}}  \right) =0 \\
\label{eq:evolution_eqs_7}
& \frac{4 u^4_{0} \mathrm{w}^2}{3 g_{1} g_{2} F \mathrm{w}} \left(g_{2} c^4-g_{1}F^4\right)-\frac{4 u^2_{0} }{3 g_{1} g_{2} F \mathrm{w}} \left(g_{2} c^2-g_{1}F^2\right) \nonumber \\ \nonumber &
+ \frac{4 g u^4_{0} c^2 F \;\Gamma_{1}}{g_{1}g_{2} \mathrm{w}} \left(d-b\right) = \Omega^2 \left\{ \frac{2(\pi^2-6)c^2 \mathrm{w}}{9 g_{1} F}
\right. \\  \nonumber  & \left.
+\frac{2 g \pi^2 u^2_{0}  c^2 d F w}{45 g_{1} g_{2}} +\frac{u^2_{0} \mathrm{w}^3}{18 g_{1} g_{2} F} \left[(\pi^2-6)c^2 \right. \right. \\  & \left. \left.
(3g_{2}c^2-gF^2)  +6(g-g_{1}) \pi^2 F^2 \right]\right\} \\ \nonumber
\label{eq:evolution_eqs_8}
& -\frac{2 u^2_{0} F}{3 g_{2} \mathrm{w}} - \frac{4 u^4_{0} \mathrm{w} F}{3 g_{1} g_{2}} \left(3 g +2 g_{1} F^2 \right)
+\frac{2 g u^4_{0} c^2 F}{g_{1} g_{2}} \left(\Gamma_{2}+2  \mathrm{w} \right) \\ \nonumber &
-\frac{2 u^2_{0} F \mathrm{w}}{g_{2}} \left(\phi^2_{1} +2 \dot{\phi_{0}} -2 b  \dot{\phi_{1}}\right) =
\frac{u^2_{0} F \mathrm{w}}{90 g_{1} g_{2}} \Omega^2  \\ & \times   \left[8 g \pi^2 b c^2 d +5\left(3\left(g_{1}-g\right)\pi^2+ 2g\left(\pi^2-6\right)c^2  \right)\mathrm{w}^2\right],
\end{align}
\end{subequations}
where $\Gamma_{1}$ and $\Gamma_{2}$ in Eq.~\eqref{eq:evolution_eqs_1}, Eq.~\eqref{eq:evolution_eqs_2} and Eq.~\eqref{eq:evolution_eqs_4} are represented as follows,
%
\begin{subequations}
\begin{align}
	\label{eq:gammas}
	&\Gamma_{1} = \mathrm{csch} \left(\frac{b-d}{\mathrm{w}}\right)^4 \left\{2 \left[2+\mathrm{cosh} \left(2\frac{b-d}{\mathrm{w}}\right)\right](b-d) \right. \\ \nonumber & \left. -3\; \mathrm{sinh} \left(2\frac{b-d}{\mathrm{w}}\right) \mathrm{w} \right\}, \\
	&\Gamma_{2} = \mathrm{csch} \left(\frac{b-d}{\mathrm{w}}\right)^2 \left\{4 \;\mathrm{cosh} \left(\frac{b-d}{\mathrm{w}}\right)(b-d) \right. \\ \nonumber & \left. - \left[ 3+ \mathrm{cosh} \left(2\frac{b-d}{\mathrm{w}}\right) \right] \mathrm{w} \right\}.
	\end{align}
	\end{subequations}
In Eqs.~\eqref{eq:evolution_eqs} we have an algebraic equation, Eq.~\eqref{eq:evolution_eqs_7}, where we do not have any derivatives of the variational parameters.  In addition we use the the constraint, Eq.~\eqref{eq:Ac}. In this case, we expect to find only 6 frequencies out of the total 8 equations of the system in Eqs.~\eqref{eq:evolution_eqs}.

\subsection{Normal modes}

\begin{figure}
	\centering
	\includegraphics[width=\columnwidth]{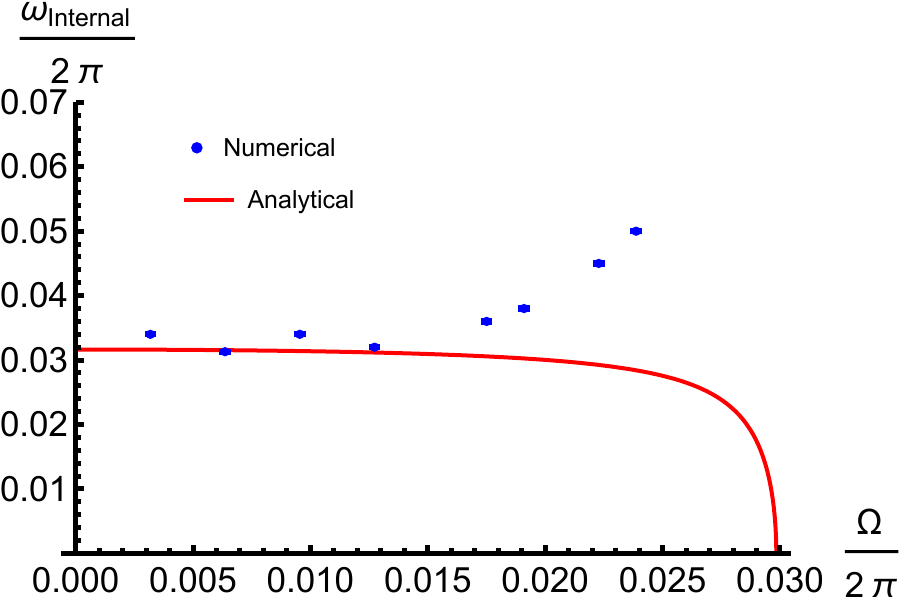}
	\caption[\emph{Internal oscillation frequency of the dark-bright soliton verses the trap frequency.}]{\emph{Internal oscillation frequency of the dark-bright soliton verses the trap frequency.} The relative degree of freedom of a dark-bright soliton is nearly independent of the center of mass degree of freedom up to a trapping frequency of about 0.0159, in units of the transverse trap frequency, at which point the internal and external motion becomes strongly coupled.  This corresponds to a trapping length ratio of $\omega^2 = 0.32$, or an approximately 3:1 prolate trap. Here we take $g_{1}=2$, $g_{2}=2.6$, $g=2.6$, $N_1/N_2=0.503$. The error bars for the numerical calculations are smaller than the point size, e.g. $\pm 0.00017$ for $\Omega/2\pi=0.0222$. }
	\label{fig:FRHPRA:paper3_plot2}
\end{figure}

The system of equations, Eqs.~\eqref{eq:evolution_eqs} has a fixed point,
\begin{align}
	\label{eq:fix_point}
	& b_{\mathrm{fp}} = 0, \; d_{\mathrm{fp}} = 0, \; A_{\mathrm{fp}} = 0, \; c_{\mathrm{fp}} = 1, \; F_{\mathrm{fp}} = 1, \\ \nonumber
	\; & w_{\mathrm{fp}} = w_{\mathrm{fp}} , \; \phi_{1fp} = 0, \; \phi_{0fp} = 0,
\end{align}
where $w_{\mathrm{fp}}$ is determined by the constraints of Eqs.~\eqref{eq:relation1}. We continue by linearizing Eqs.~\eqref{eq:evolution_eqs} around the fixed point Eq.~\eqref{eq:fix_point}. Here we set,
\begin{align}
	\label{eq:expand_around_fix_point}
	& a_{j}\left(t\right) = a_{j\mathrm{fp}} + \delta a_j \; e^{i \omega t},
\end{align}
where $\omega$ is the oscillation frequency between the two components and the $a_{j}$ are the 8 variational parameters. Keeping $\delta a_j$ to linear order results in a matrix equation of the form,
 \begin{equation}
	\label{eq:matrix1}
\begin{bmatrix}
    A_{11} & A_{12} & A_{13} & 0 & 0 & 0 & 0 & 0 \\
	0 & A_{22} & A_{23} & A_{24} & 0 & 0 & 0 & 0 \\
	0 & A_{32} & 0 & A_{34} & 0 & 0 & 0 & 0 \\
	0 & 0 & 0 & 0 & A_{45} & A_{46} & A_{47} & 0 \\
	0 & 0 & 0 & 0 & A_{55} & A_{56} & 0 & 0 \\
	A_{61} & 0 & A_{63} & 0 & A_{65} & A_{66} & 0 & 0 \\
	0 & 0 & 0 & 0 & A_{75} & A_{76} & A_{77} & 0 \\
	0 & 0 & 0 & 0 & A_{85} & A_{86} & A_{87} & A_{88} \\
\end{bmatrix}
\begin{bmatrix}
    \delta \phi_{1}   \\
    \delta d   \\
    \delta b   \\
    \delta A   \\
    \delta F   \\
    \delta \mathrm{w}   \\
    \delta c   \\
    \delta \phi_{0}   \\
\end{bmatrix}
=\begin{bmatrix}
	0
\end{bmatrix}
\end{equation}
where $[0]$ refers to a column vector with eight entries of value zero. The nonzero terms are written in Appendix~\ref{appendix:Matrix elements}.
%
%
Taking the determinant of the matrix and solving for the eigenfrequencies, $\omega$, we obtain,

\begin{align}
	\label{eq:omega}
	\alpha_{1} \omega^6 + \alpha_{2} \omega^4 + \alpha_{3} \omega^2 =0\,,
\end{align}
where as mentioned already only six eigenfrequencies are expected due to constraints and the form of the coupled nonlinear ODEs in Eqs.~\ref{eq:evolution_eqs}.  Solving the determinant we obtain,
%
\begin{align}
	\label{eq:frequency}
	\omega_{\pm} & = 0, \; 0, \;  \frac{1}{\sqrt{2}} \sqrt{-\frac{\alpha_{2}}{\alpha_{1}} \pm \frac{1}{\alpha_{1}} \sqrt{\alpha^2_{2} - 4 \alpha_{1} \alpha_{3}}}, \; \\ \nonumber &
	-\frac{1}{\sqrt{2}} \sqrt{-\frac{\alpha_{2}}{\alpha_{1}} \pm \frac{1}{\alpha_{1}} \sqrt{\alpha^2_{2} - 4 \alpha_{1} \alpha_{3}}}.
\end{align}
where we write out the long expressions for $\alpha_{1}$, $\alpha_{2}$ and $\alpha_{3}$ in Appendix~\ref{appendix:Matrix elements}. Since we are considering a small oscillation frequency, $\Omega \ll 1$, we expand the coefficients (i.e., $\alpha_{1}, \alpha_{2}$ and $\alpha_{3}$) around $\Omega \to 0$ and find that $\alpha_{3} \to 0$. Therefore, we end up with one internal oscillation frequency of interest,
\begin{align}
	\label{eq:frequency_one}
	\omega_{\mathrm{internal}} & = \sqrt{\frac{\alpha_{2}}{-\alpha_{1}}}.
\end{align}
The dark-bright soliton we consider exists in repulsive media, therefore, $g$, $g_{1}$ and $g_{2}$ all take positive values. In this case, $\alpha_{1} < 0$, $\alpha_{2}>0$ for any values of the free parameters mentioned in Sec.~\ref{sec:FRHPRA:Lagrangian density and ansatz}. In Fig.~\ref{fig:FRHPRA:paper3_plot2} we plot a typical case for in the internal oscillation frequency, Eq.~\ref{eq:frequency_one}, using the same parameters as our previous treatment of the uniform case for comparison~\cite{majed2017}.  The result is nearly independent of trapping frequency until a sudden strong coupling for larger $\Omega$, beyond which the result turns imaginary.  However, this is also beyond the assumptions of the model, namely $\Omega \ll 1$.  Therefore we examine the questions of the real trend in a more thorough numerical treatment in Sec.~\ref{sec:FRHPRA:Numerical calculations}.

\subsection{Nonlinear dark-bright soliton motion}
\label{sec:FRHPRA:dark-bright_soliton_motion}

\begin{figure}
	\centering
	\includegraphics[width=\columnwidth]{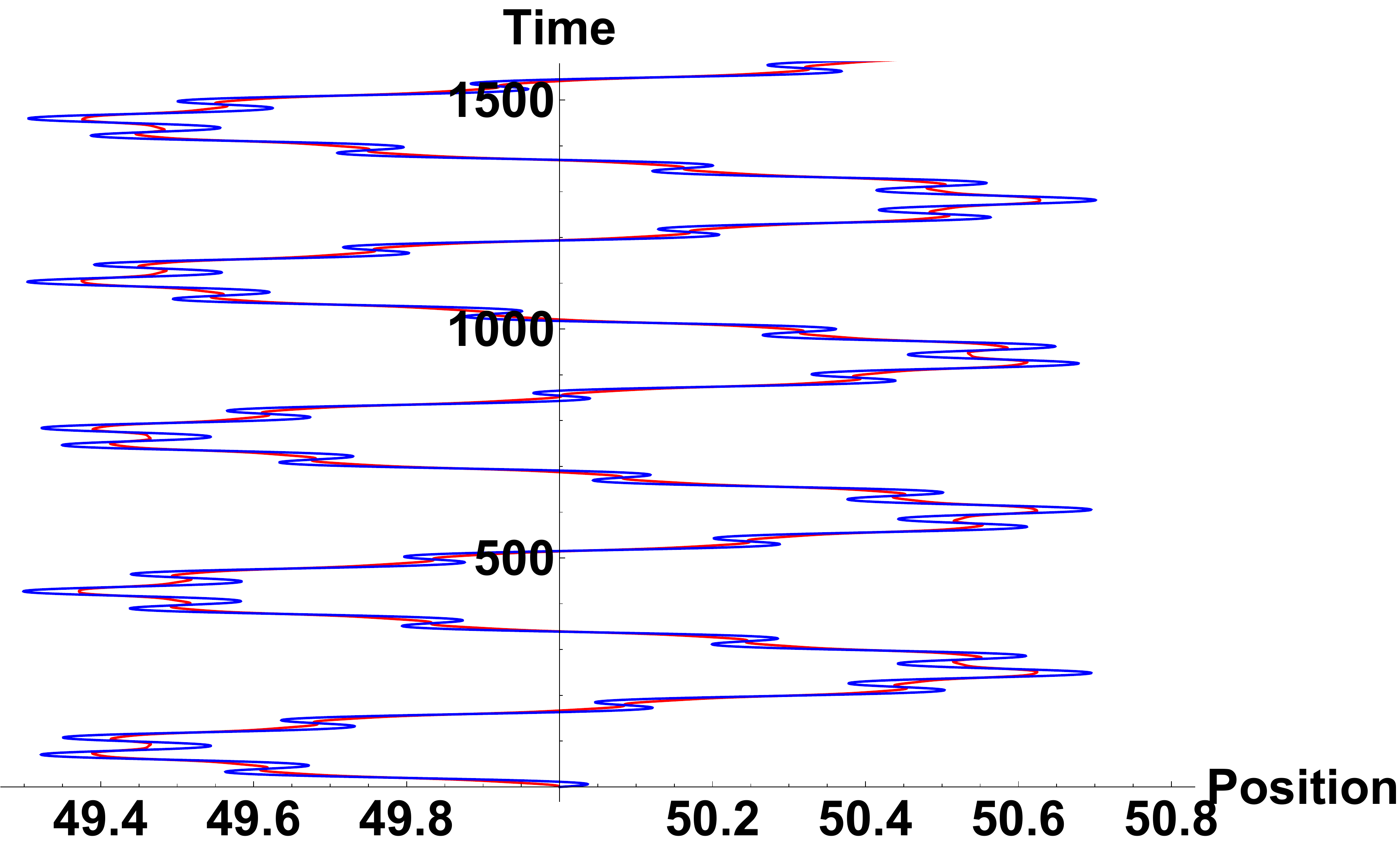}
	\caption[\emph{Oscillation of dark-bright soliton in a harmonic potential well.}]{\emph{Oscillation of dark-bright soliton in a harmonic potential well.} The nonlinear ODE evolution of the dark and bright soliton positions resulting from our variational Lagrangian treatment shows a rich structure to the internal dynamics, even for a small trapping frequency of $\Omega/2\pi=0.0064$.  The free parameters are the same as in Fig.~\ref{fig:FRHPRA:paper3_plot2}.}
	\label{fig:FRHPRA:analy_osci_1}
\end{figure}

The system of Eqs.~\eqref{eq:evolution_eqs} also can be simplified to a smaller set of second order nonlinear coupled ODEs. From Eq.~\eqref{eq:evolution_eqs_5} and Eq.~\eqref{eq:evolution_eqs_6}, we obtain the following,
\begin{equation}
	\dot{b} = - \phi_{1},
\end{equation}
with the help of Eq.~\eqref{eq:evolution_eqs_1}, we get our first second order differential equation (ODE),
%
	\begin{align}
		\label{eq:2nd_ODE_b}
	\ddot{b} = &-\frac{g u^2_{0} c^2}{g_{1} \mathrm{w}} \Gamma_{1} -\frac{\Omega^2}{45 g_{1}} \\ \nonumber
	\times & \left\{ 45(g_{1}-g)  -g(\pi^2-15)c^2+g \pi^2 c^2 d\right\} b.
	\end{align}
%
Note that when we set $c=0$ (i.e., eliminating the dark soliton) Eqs.~\eqref{eq:2nd_ODE_b} recovers the well-known oscillation frequency of the one-component bright soliton in a harmonic potential,
	\begin{equation}
		\label{eq:2nd_ODE_b_without_DS}
	\ddot{b} + \Omega^2 b =0.
	\end{equation}
In the limiting case, $g=0$ because there is no interaction between the bright soliton and the dark soliton. The second ODE is obtain by inserting Eq.~\eqref{eq:evolution_eqs_3} into Eq.~\eqref{eq:evolution_eqs_4} and use the normalization conditions, Eqs.~\eqref{eq:relation1}, we obtain,
\begin{align}
		\label{eq:1sr_ODE_d_new}
	\dot{d} &= \frac{1}{576 A (1-A)^{5/2}} \left[\frac{3 g^3_{1} N^3_{1} \pi^2 \Omega^2}{N^{3} u^6_{0}}  \right. \\ \nonumber & \left.
	+\frac{4 g^3_{1} N^3_{1}(3+\pi^2)\Omega^2 (-1+A^2)}{N^3 u^6_{0}} \right. \\ \nonumber & \left.  -\frac{72 g_{1} N_{1} \Omega^2 (-1+A^2)^{2}}{N u^4_{0}} \right. \\ \nonumber & \left.
	+ \frac{96(2 g_{1} N_{1} - 3 g N_{2})(-1+A^{2})^{3}}{N}  \right. \\ \nonumber & \left.
	- \frac{96 u^2_{0} (4 N - 3 g N_{2} \Gamma_{2})(-1+A^{2})^{4}}{g_{1} N_{1}}.
	 \right]
\end{align}
Equation~\eqref{eq:1sr_ODE_d_new} take the form $\dot{d} = f(A(t),\Gamma_2(t))$. Taking the total time derivative of Eq.~\eqref{eq:1sr_ODE_d_new} yields,
\begin{equation}
	\label{eq:2nd_ODE_d}
	\ddot{d} = \alpha \dot{A},
\end{equation}
where $\alpha$ is obtained from Eq.~\eqref{eq:1sr_ODE_d_new} and Eq.~\eqref{eq:evolution_eqs_2},
 \begin{align}
 	\label{eq:2nd_ODE_alpha}
 	\alpha &= -\frac{g N_{2} \Gamma_{1}}{6 g^3_{1} N^3_{1}} \left(-4-4 g^2_{1} N^2_{1} + 6 g g_{1} N_{1} N_{2} +3 g N_{2} \Gamma_{2}  \right) \\ \nonumber
	& + \frac{\Gamma_{1}}{6 g^3_{1} N^3_{1} A^2} \left(-4 g N_{2} -2 g g^2_{1} N^2_{1} N_{2} + 3 g^2 g_{1} N_{1} N^2_{2}  \right. \\ \nonumber & \left.
	+ 3 g^2 N^2_{2} \Gamma_{2}\right) + \Omega^2 \left\{ \frac{g N_{2} \Gamma_{1}}{576 g_{1} N_{1} A^2} \left[-72-12 g^2_{1} N^2_{1} \right. \right. \\ \nonumber & \left. \left.  -g^2_{1} N^2_{1} \pi^2 +216 A^2 +48 g^2_{1} N^2_{1} A^2 + 7 g^2_{1} N^2_{1} \pi^2 A^2 \right]   \right. \\ \nonumber & \left.
	- \frac{1}{72 g^2_{1} N^2_{1} A^2} \left[32+20 g^2_{1} N^2_{1} +2 g^4_{1} N^4_{1}  \right. \right. \\ \nonumber & \left. \left.
	-3g g^3_{1} N^3_{1} N2 -24 g N_{2} \Gamma_{2} -12 g^2_{1} N^2_{1} A^2 \right. \right. \\ \nonumber & \left. \left.
	+ 24 g g_{1} N_{1} N_{2} \left(-1+A^2\right)  -3 g g^2_{1} N^2_{1} N_{2} \Gamma_{2} \left(1+A^2\right)  \right] d.
\right\}
 \end{align}
By plotting Eq.~\eqref{eq:2nd_ODE_b} and Eq.~\eqref{eq:2nd_ODE_d} we obtain Fig.~\ref{fig:FRHPRA:analy_osci_1}, where the interplay between external and internal degrees of freedom of the dark-bright soliton is clearly evident, showing that the assumption of the two components moving together, as found in previous treatments before this Article, does not capture the richness of the dynamics.

%
%
%

\section{Full numerical evolution of the coupled GPEs}
\label{sec:FRHPRA:Numerical calculations}


We now numerically study the oscillation of the dark-bright and the internal oscillation between the two components in a harmonic potential described by Eq.~\eqref{eq:potential1}, making no other assumptions beyond coupled GPEs. Throughout this section, we present the simulations with grid size $n_{x}$ = 256 in a box with hard-wall boundaries, noting that this is sufficient to converge our simulations.  For example, the error bars are smaller than the point size for internal frequencies (see Fig.~\ref{fig:FRHPRA:paper3_plot2}) even when we cut the grid in half to 128 points.  The box length is set to $L$ = 100 unless otherwise noted.

\begin{figure}
\begin{subfigure}
  \centering
  \includegraphics[width=0.9\columnwidth]{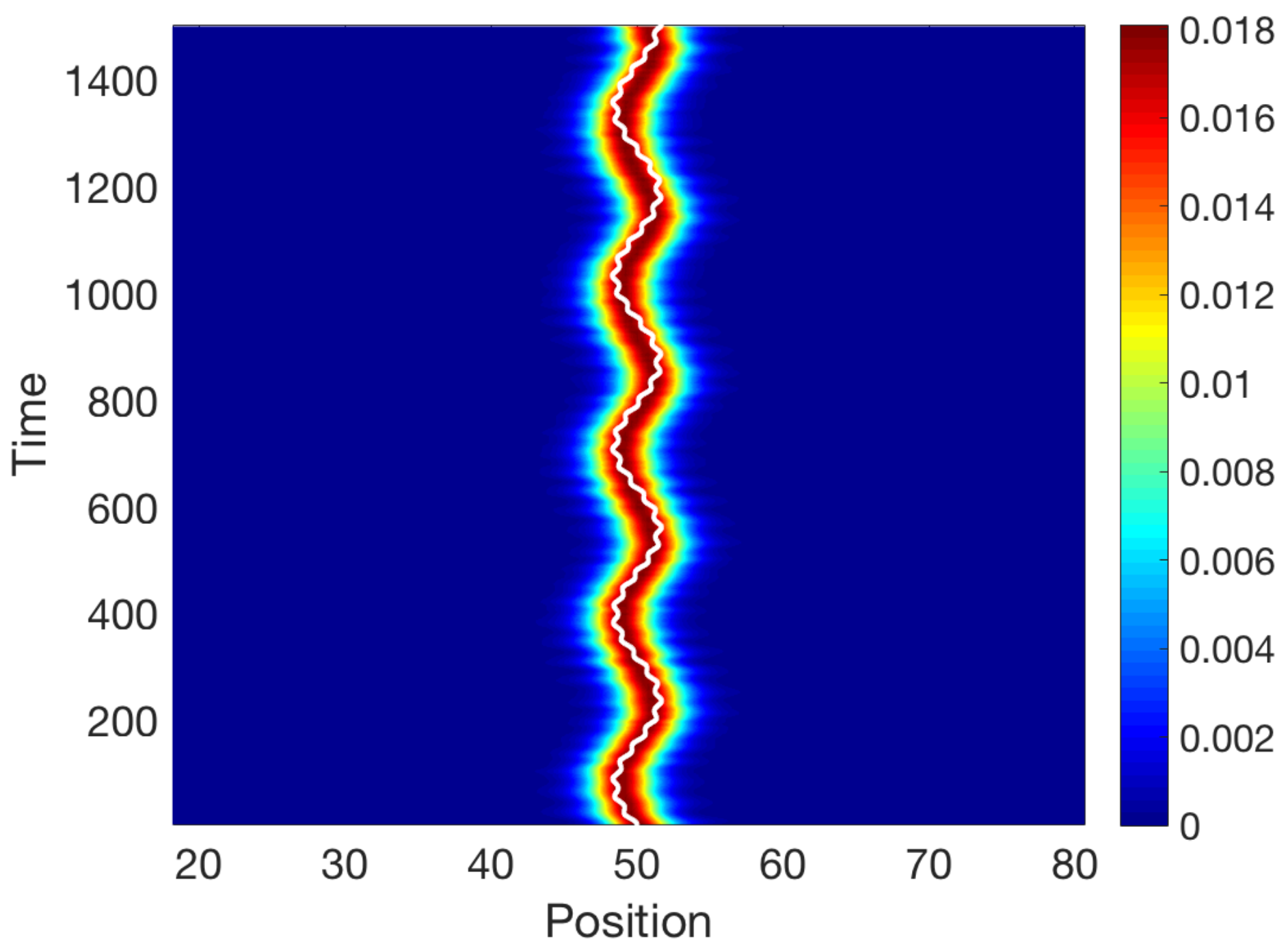}
\end{subfigure}%
\begin{subfigure}
  \centering
  \includegraphics[width=0.9\columnwidth]{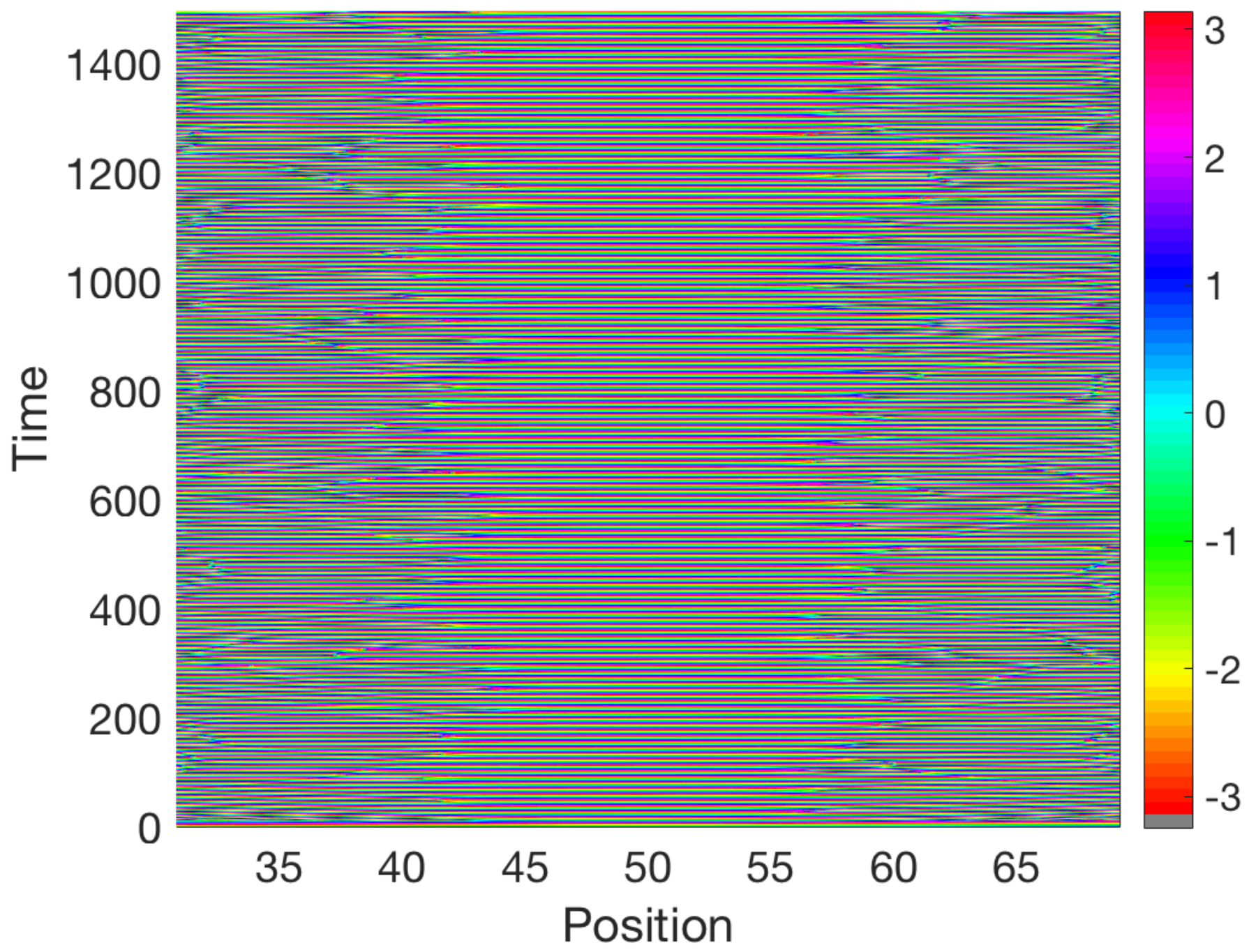}
\end{subfigure}
\caption[\emph{Bright component in dark-bright soliton}.]{\emph{Bright component in dark-bright soliton.} The oscillation of the bright soliton component in dark-bright soliton. The white line represents the analytical result for the bright soliton position, Eq.~\eqref{eq:2nd_ODE_b}. We set the trap frequency $\Omega/2\pi = 0.0064$. We find the dark-bright soliton oscillates with $\omega_{\mathrm{DB}} / 2 \pi  = 0.0039$. In the lower panel, we plot the phase.}
\label{fig:FRHPRA:bright_component_in_DB_num}
\end{figure}

\begin{figure}
\begin{subfigure}
  \centering
  \includegraphics[width=0.9\columnwidth]{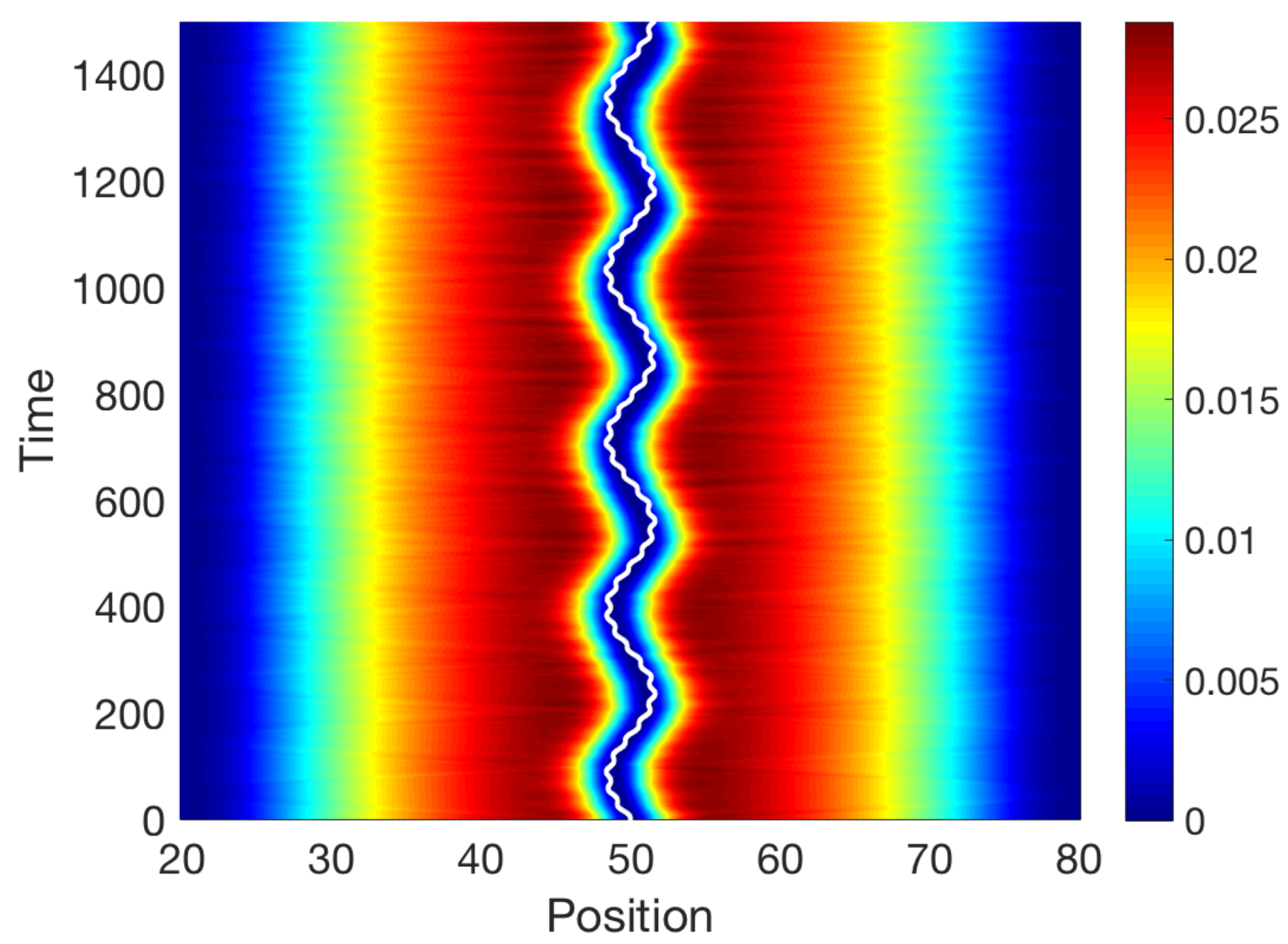}
\end{subfigure}%
\begin{subfigure}
  \centering
  \includegraphics[width=0.9\columnwidth]{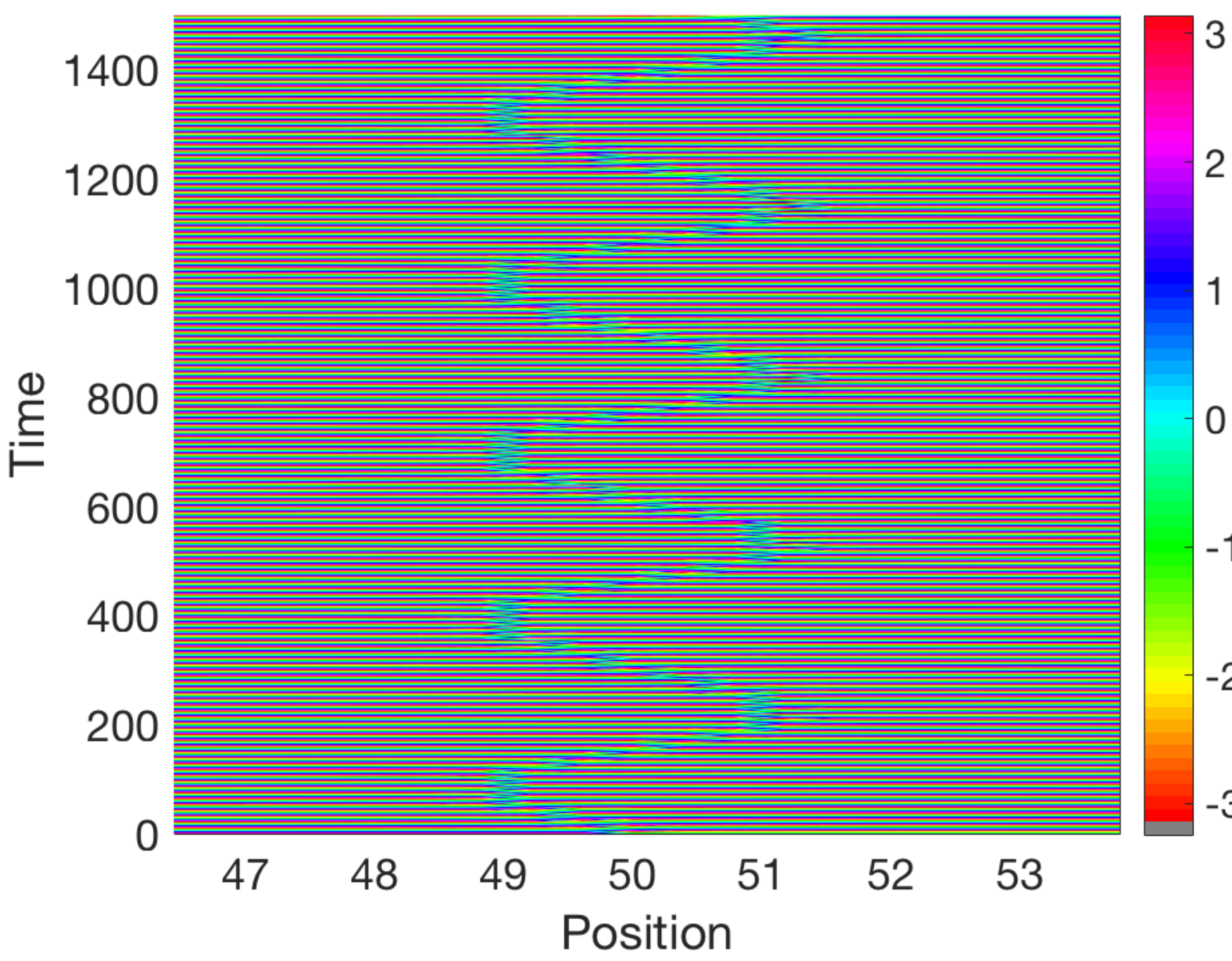}
\end{subfigure}
\caption[\emph{Dark component in dark-bright soliton}.]{\emph{Dark component in dark-bright soliton.} The oscillation of the dark soliton component in dark-bright soliton. In the upper panel, the white line represents the analytical plot from Eq.~\eqref{eq:2nd_ODE_b}. We set the trap frequency $\Omega/2\pi = 0.0064$. We find the dark-bright soliton oscillates with $\omega_{\mathrm{DB}}/ 2 \pi = 0.0039$. In the lower panel, we plot the phase. }
\label{fig:FRHPRA:dark_component_in_DB_num}
\end{figure}


\subsection{Dark-bright soliton in harmonic potential}
\label{sec:FRHPRA:Dark-bright soliton in harmonic potential}
To move a dark-bright soliton in a harmonic potential, we may imprint a phase on the bright component or the dark component but with a fundamental difference between these two methods. If we imprint a phase difference on the dark component only, it will move slowly such that it will pull the bright component with it but without any oscillation between the two components. For this method, it is noteworthy to mention that an ansatz with only one variable to represent the location of the dark and bright components is a valid choice to describe the moving dark-bright in a harmonic potential as this is the case for other studies~\cite{PhysRevA.84.053626}. But since we are interested in the oscillation of dark-bright soliton in a harmonic potential with an additional degree of freedom, namely, the internal oscillation of the two components, we work with the second method (i.e., imprinting a phase on the bright component only). In this method, the relatively small density of the bright component moves faster when imprinting a phase on it and as a result, it will drag the dark soliton with it and form an oscillation between the two components. Therefore, the dark-bright soliton will move, and we study the center-of-mass trajectory to calculate the oscillation of the dark-bright soliton as a whole.

In Fig.~\ref{fig:FRHPRA:bright_component_in_DB_num} and Fig.~\ref{fig:FRHPRA:dark_component_in_DB_num}, we plot the outcomes from the numerical simulations and the analytical calculations of the bright and dark components, respectively. In each plot, the upper panel shows the density, and the lower panel the phase. The analytical results, the white line in the center of the bright and dark components, oscillate with nearly the same frequency as the numerical results, showing a small deviation after many trap periods.  This deviation is a result of the interaction between the dark-bright soliton with the reflected phonons, not captured in the analytical model where we assumed an inert Thomas-Fermi background. When the dark-bright soliton moves in a harmonic potential, phonons are created and propagate away with the speed of sound. They then reach the low density regions of the BEC at the harmonic trap edges and turn back around to interact with the dark-bright soliton.

\begin{figure}
	\centering
	\includegraphics[width=\columnwidth]{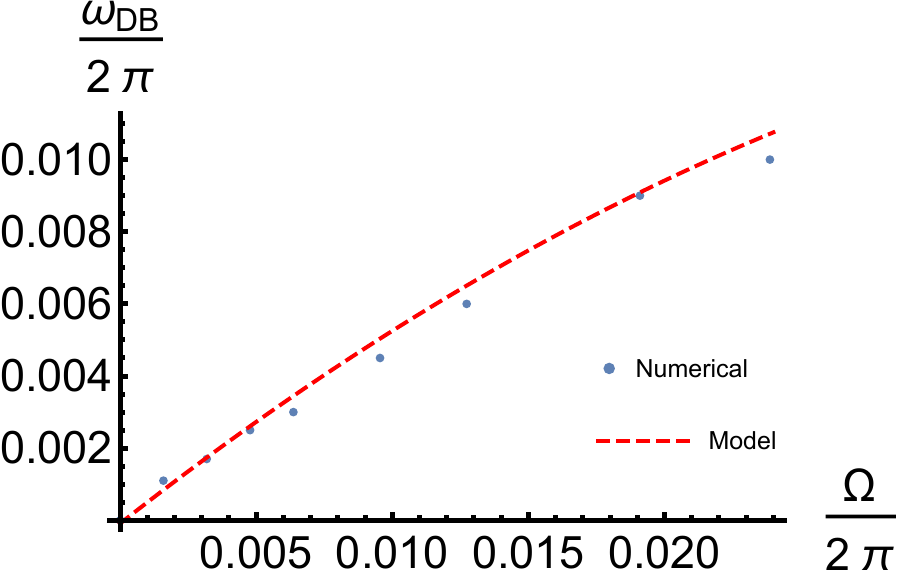}
	\caption{  \emph{The oscillation of the dark-bright soliton for different values of the trap oscillation of the harmonic potential.} We compare the analytical predictions to numerical results of the oscillation of dark-bright soliton, $\omega_{\mathrm{DB}}/2\pi$, for a wide range of trap frequencies, $\Omega/2\pi$.}
	\label{fig:FRHPRA:paper3_plot1}
\end{figure}

To test the analytical predictions against the numerical outcomes, we plot the center of mass oscillation frequency $\omega_\mathrm{DB}$ of the dark-bright soliton vs. the trapping frequency $\Omega$ in Fig.~\ref{fig:FRHPRA:paper3_plot1}. The analytical results are obtained by evolving the nonlinear ODEs and performing a Fourier transform.  These scale almost linearly together showing they are nearly but not quite proportional for weak trapping.  For small trapping frequencies, as shown in Fig.~\ref{fig:FRHPRA:paper3_plot2}, the internal frequency is also nearly independent of the trap.   This is an indication that the internal oscillation of the two components does not couple with the oscillation of the dark-bright soliton in the weak trapping case.  In contrast, our coupled GPE simulations show that for stronger trapping the internal degree of freedom is strongly dependent on the trap frequency, see Fig.~\ref{fig:FRHPRA:paper3_plot2}.  In this regime, the analytical result diverges to zero, but the numerical result increases.  We interpret these results further in Sec.~\ref{sec:FRHPRA:Conclusions}.
%

\begin{figure}
\begin{subfigure}
  \centering
  \includegraphics[width=\columnwidth]{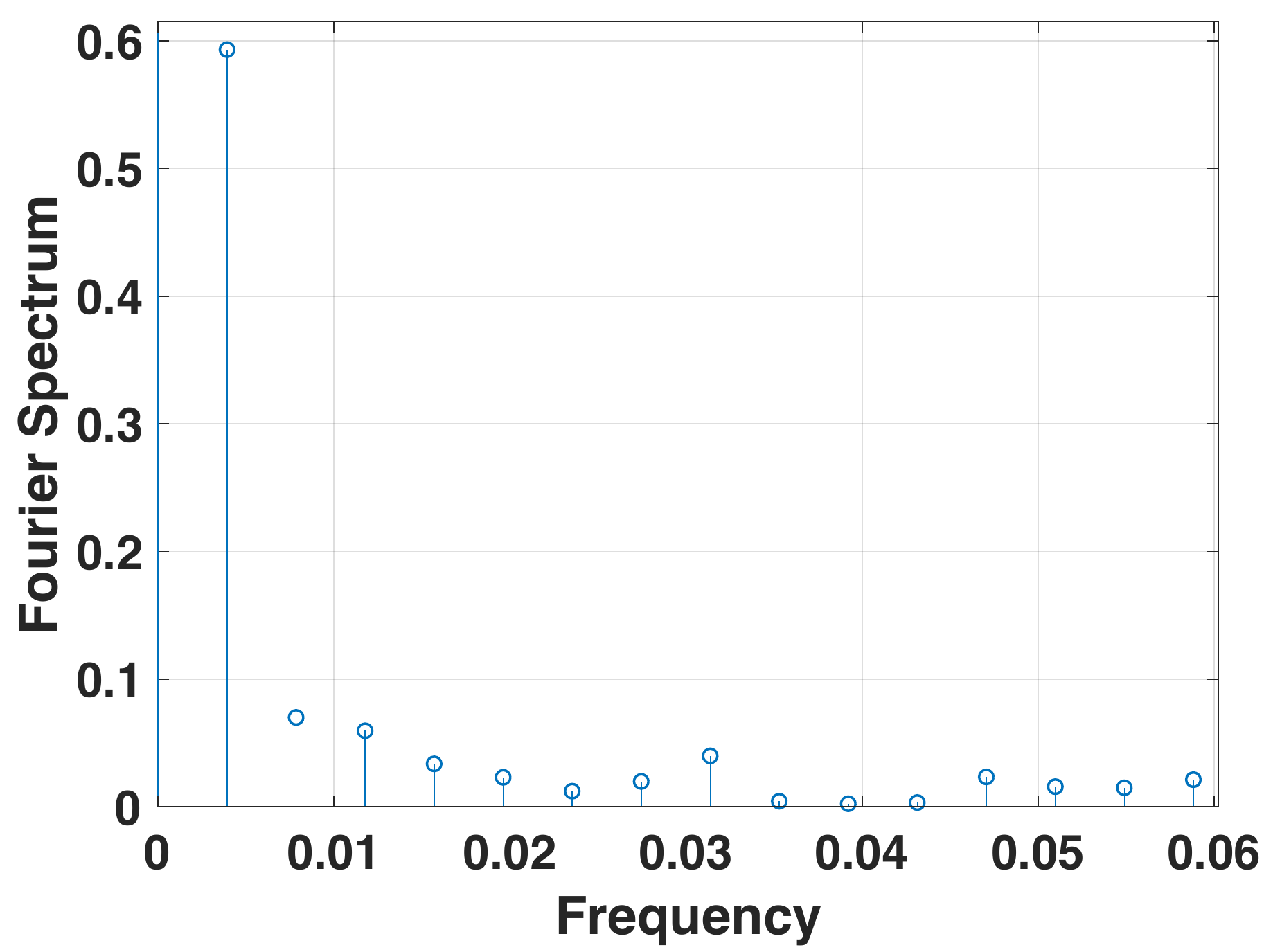}
\end{subfigure}%
\begin{subfigure}
  \centering
  \includegraphics[width=\columnwidth]{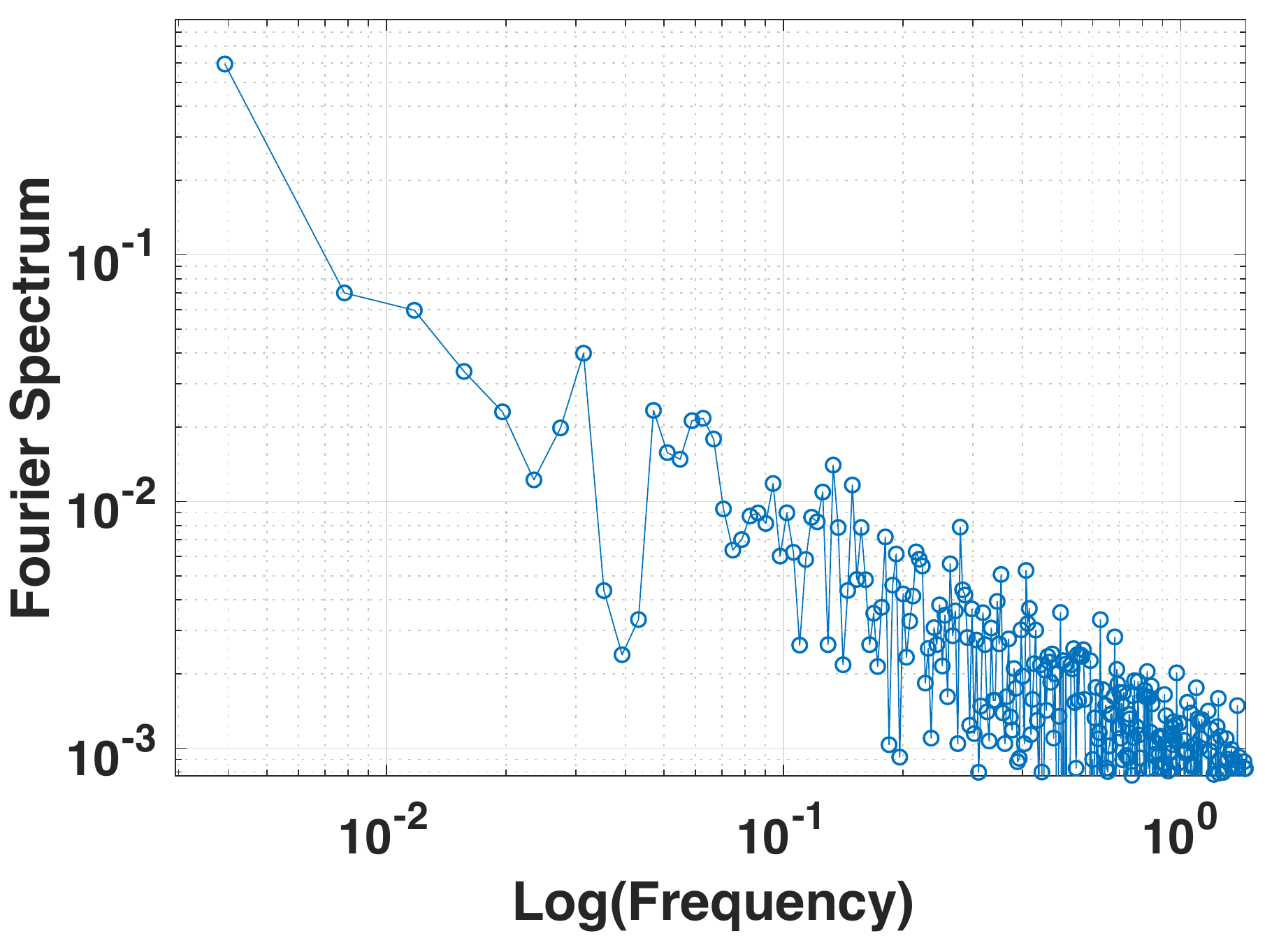}
\end{subfigure}
\caption[\emph{Oscillation frequencies of the dark-bright soliton in harmonic potentail}.]{\emph{Oscillation frequencies of the dark-bright soliton in harmonic potential}. A Fourier transform of our numerical results allows us to pick out the important frequencies in the problem.  We show here a sample case of $\Omega/2\pi = 0.0064$.  The first dominant frequency is located at $\omega /2 \pi = 0.0039$  which corresponds to the center of mass oscillation of the dark-bright soliton in the harmonic potential. The second dominant frequency is located at $\omega /2 \pi = 0.032$ , and corresponds to the internal oscillation between the two components.  Overall the dynamics is in fact quite rich, with many aspects to the motion, as observed in the dense Fourier tail.}
\label{fig:FRHPRA:frequency_numerical}
\end{figure}


\subsection{Robustness of dark-bright soliton oscillations}
\label{sec:FRHPRA:Dark-bright soliton in harmonic potential with white noise}

\begin{figure}
\begin{subfigure}
  \centering
  \includegraphics[width=\columnwidth]{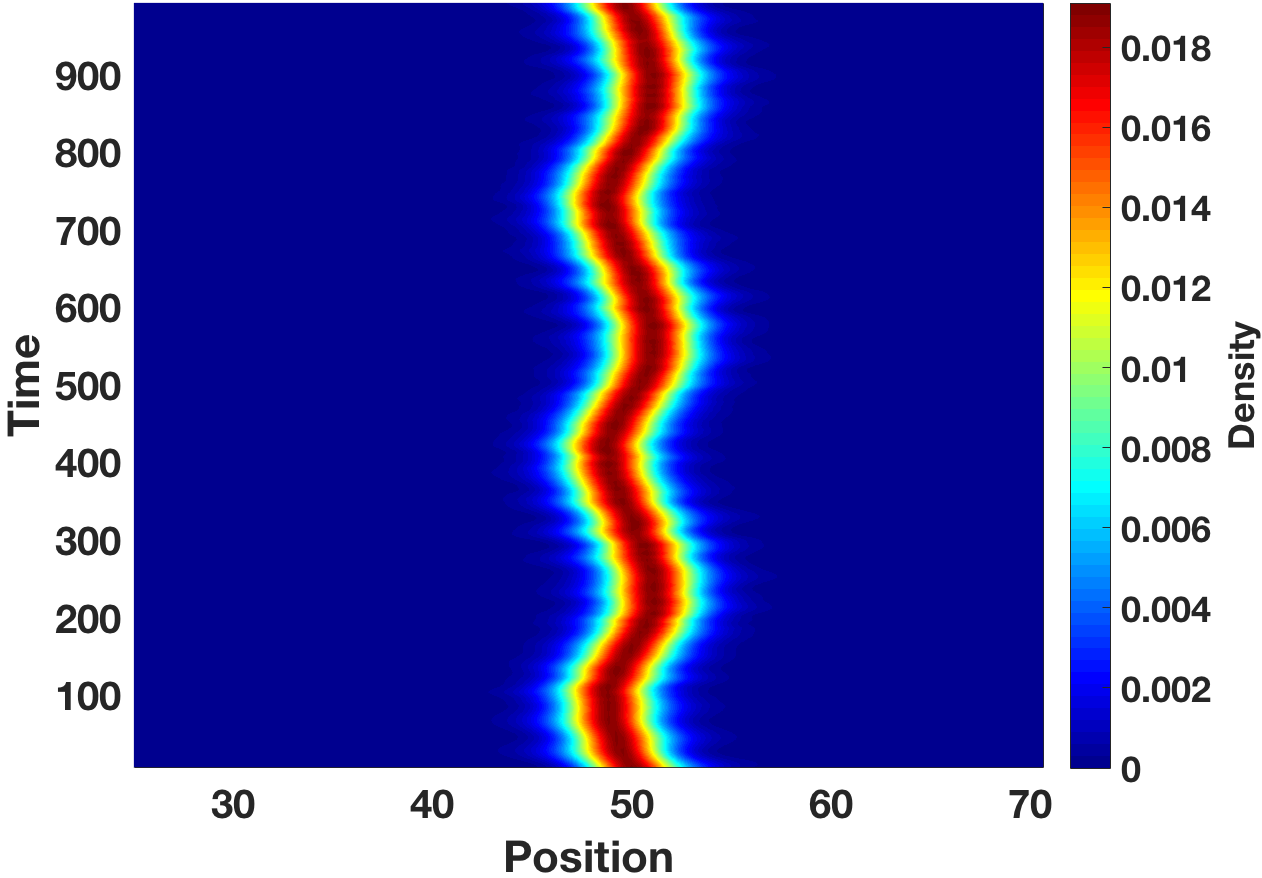}
\end{subfigure}%
\begin{subfigure}
  \centering
  \includegraphics[width=\columnwidth]{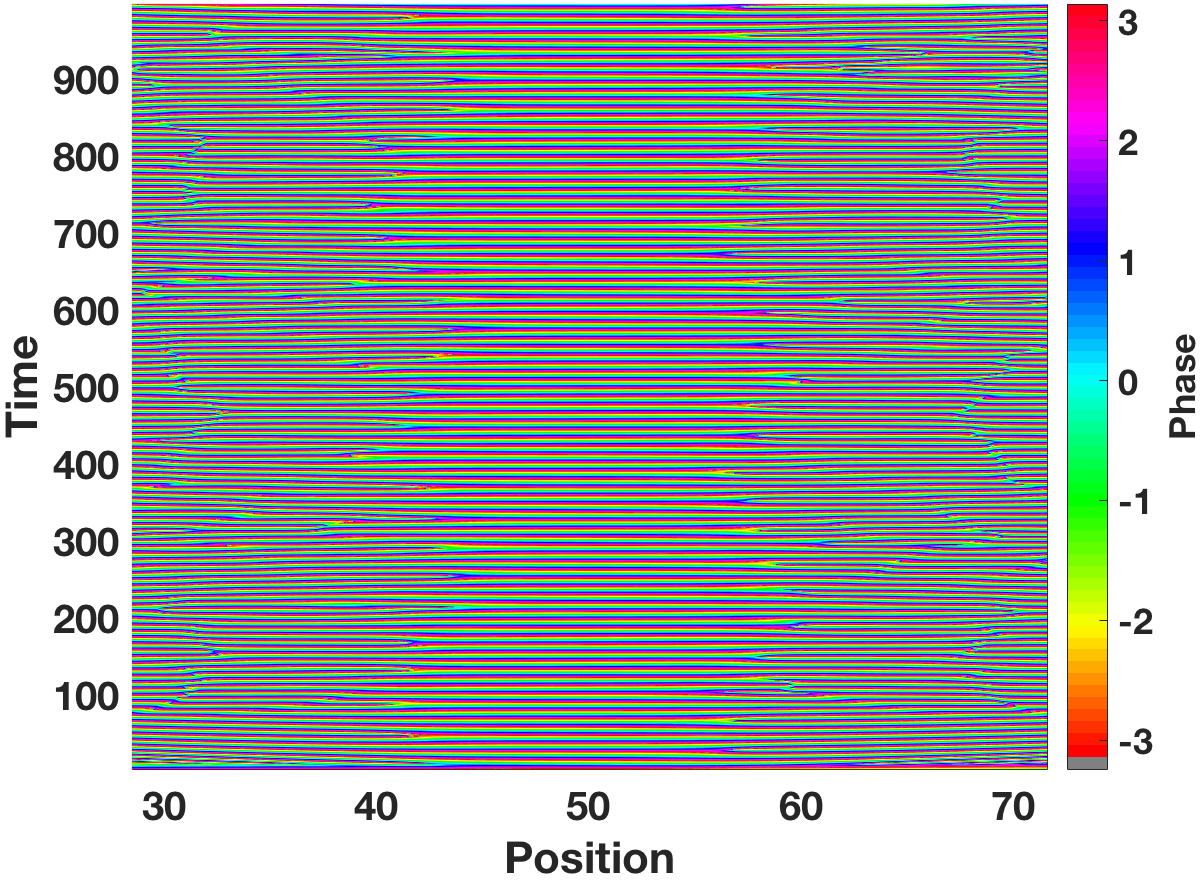}
\end{subfigure}
\caption[\emph{Oscillation of bright component in a harmonic potential}.]{\emph{Oscillation of bright component in a harmonic potential.} We plot the density (phase) in the upper (lower) panel for the bright component in dark-bright soliton with harmonic potential frequency, $\Omega/2\pi = 0.0064$, $g_{1}=2$, $g_{2}=2.6$ and $g=2.6$.}
\label{fig:FRHPRA:BS_den_no_noise}
\end{figure}

\begin{figure} 
\begin{subfigure}
  \centering
  \includegraphics[width=\columnwidth]{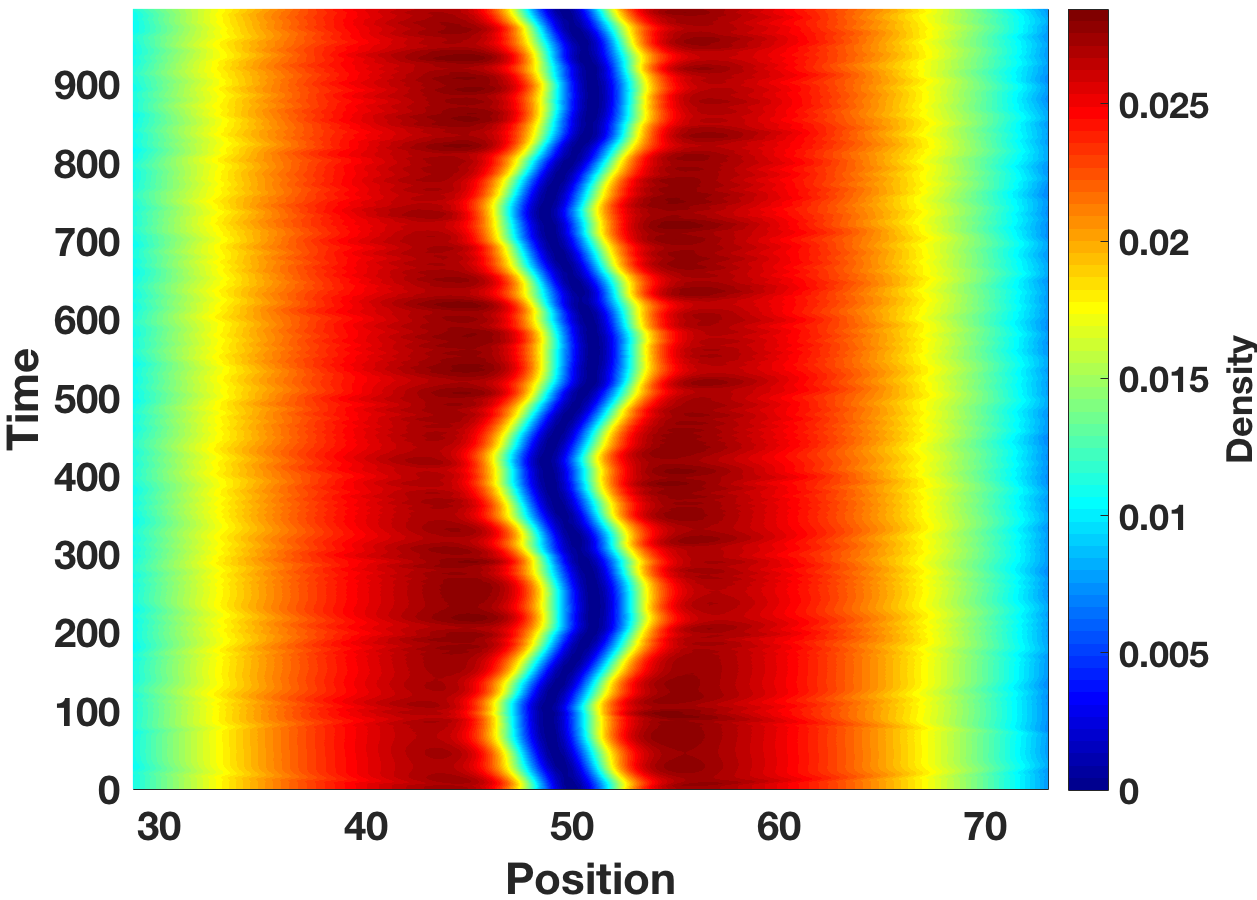}
\end{subfigure}%
\begin{subfigure}
  \centering
  \includegraphics[width=\columnwidth]{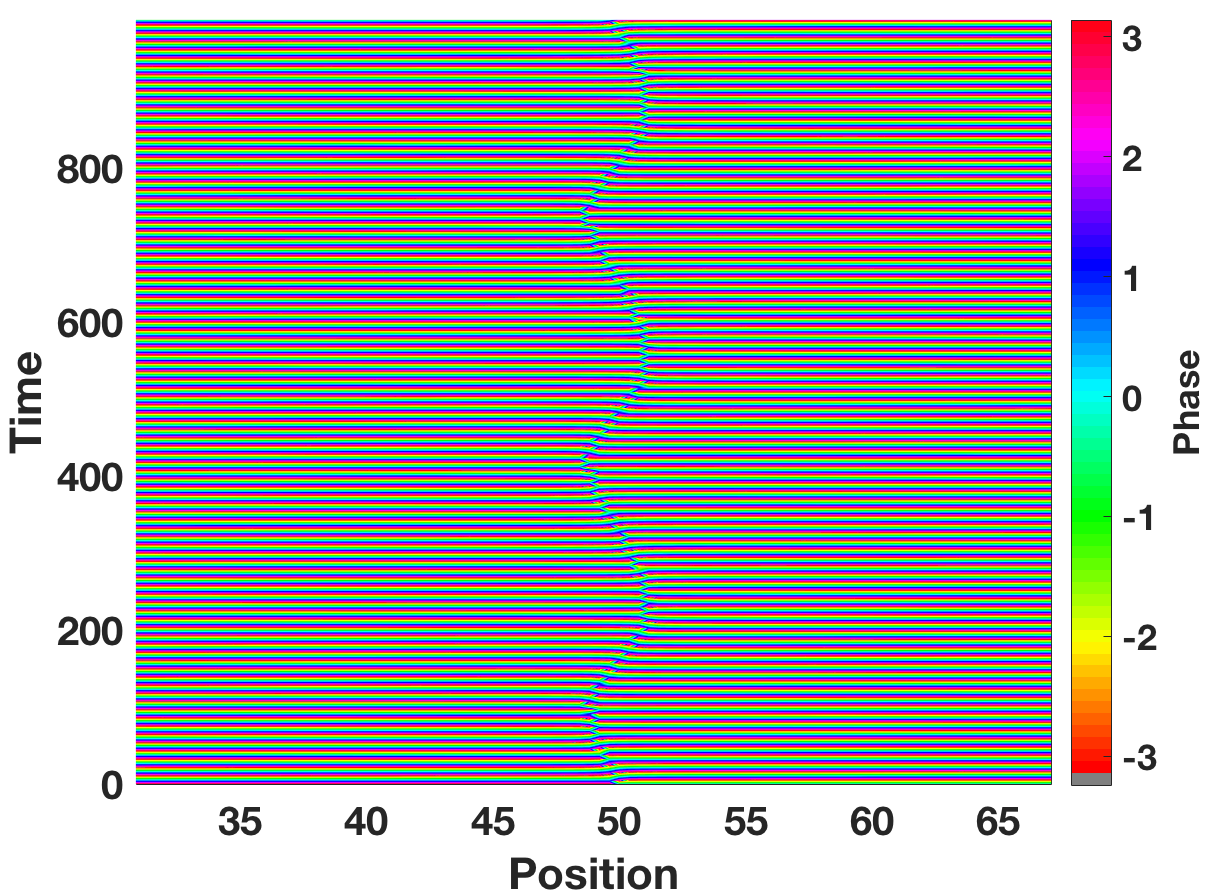}
\end{subfigure}
\caption[\emph{Oscillation of dark component in a harmonic potential}.]{\emph{Oscillation of dark component in a harmonic potential.} We plot the density (phase) in the upper (lower) panel for the dark component in dark-bright soliton with harmonic potential frequency, $\Omega/2\pi = 0.0064$, $g_{1}=2$, $g_{2}=2.6$ and $g=2.6$.}
\label{fig:FRHPRA:DS_den_no_noise}
\end{figure}

\begin{figure}
\begin{subfigure}
  \centering
  \includegraphics[width=\columnwidth]{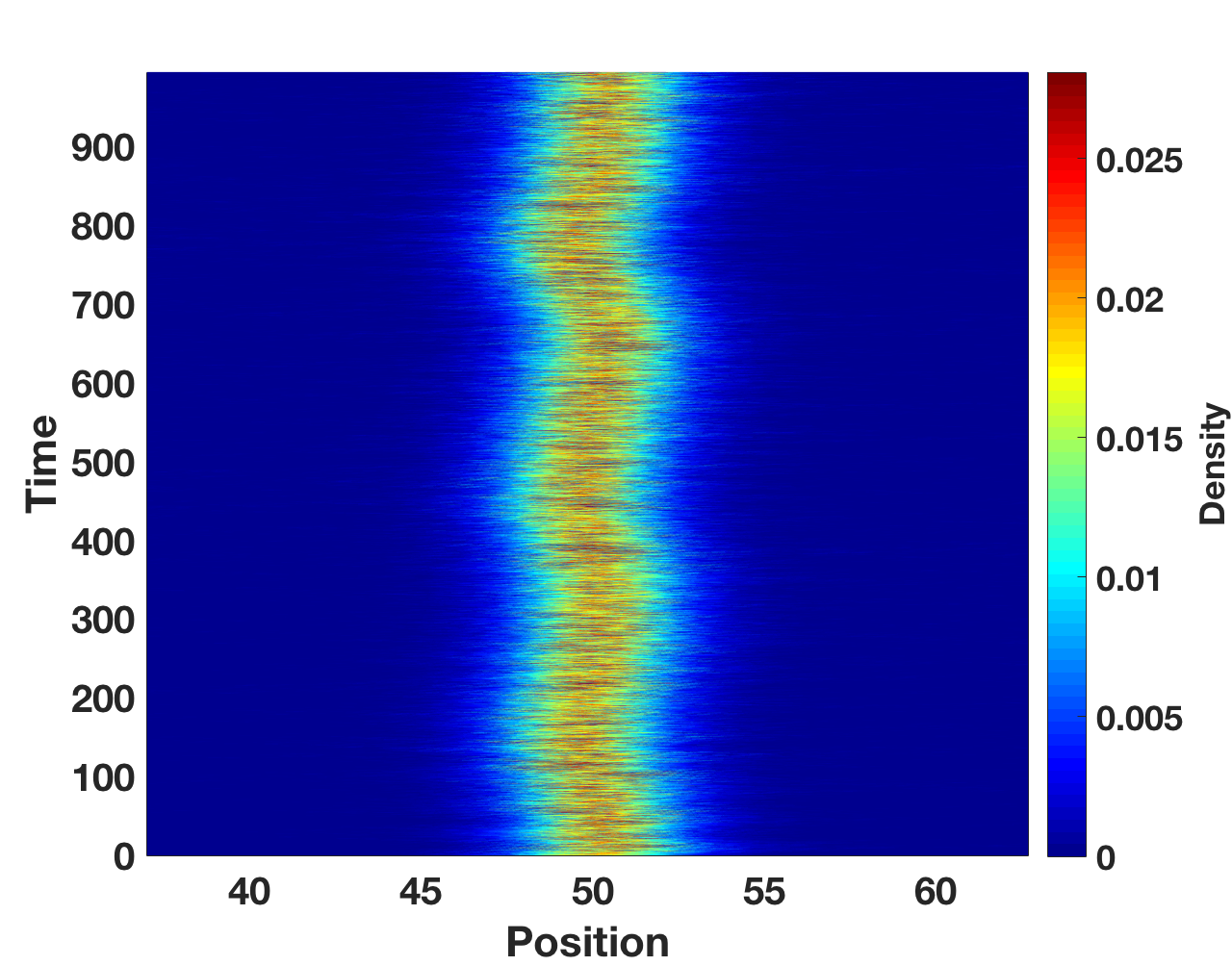}
\end{subfigure}%
\begin{subfigure}
  \centering
  \includegraphics[width=\columnwidth]{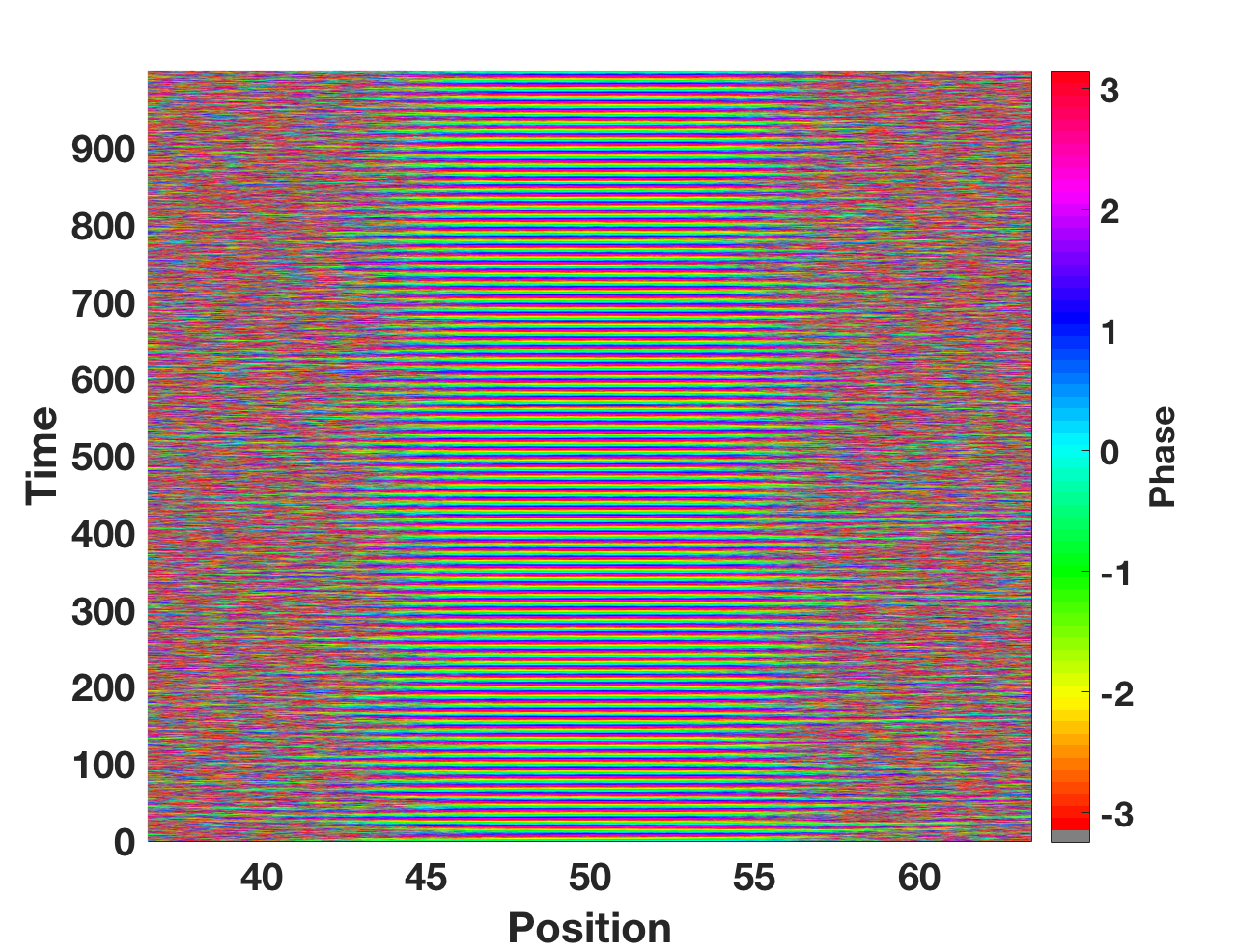}
\end{subfigure}
\caption[\emph{Oscillation of bright component in a harmonic potential when white noise added}.]{\emph{Oscillation of bright component in a harmonic potential when white noise added}. We plot the density (phase) in the upper (lower) panel for the bright component in dark-bright soliton for the same parameters in Fig.~\ref{fig:FRHPRA:BS_den_no_noise}  with $5\%$ noise added to the initial wave function at $t=0$. The bright component oscillate with the same frequency in Fig.~\ref{fig:FRHPRA:BS_den_no_noise} but with less oscillation amplitude.}
\label{fig:FRHPRA:BS_den_with_noise}
\end{figure}

\begin{figure}
\begin{subfigure}
  \centering
  \includegraphics[width=\columnwidth]{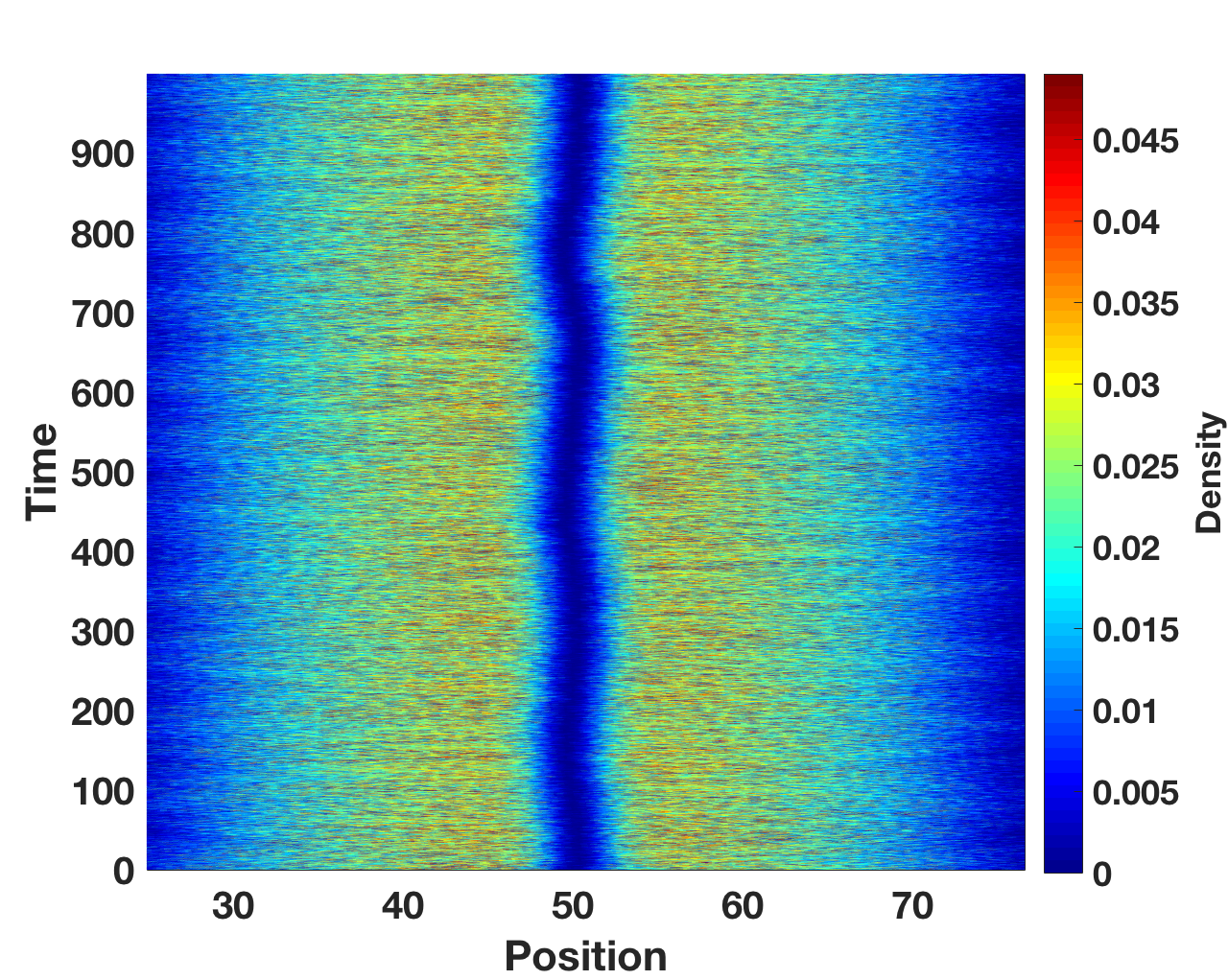}
\end{subfigure}%
\begin{subfigure}
  \centering
  \includegraphics[width=\columnwidth]{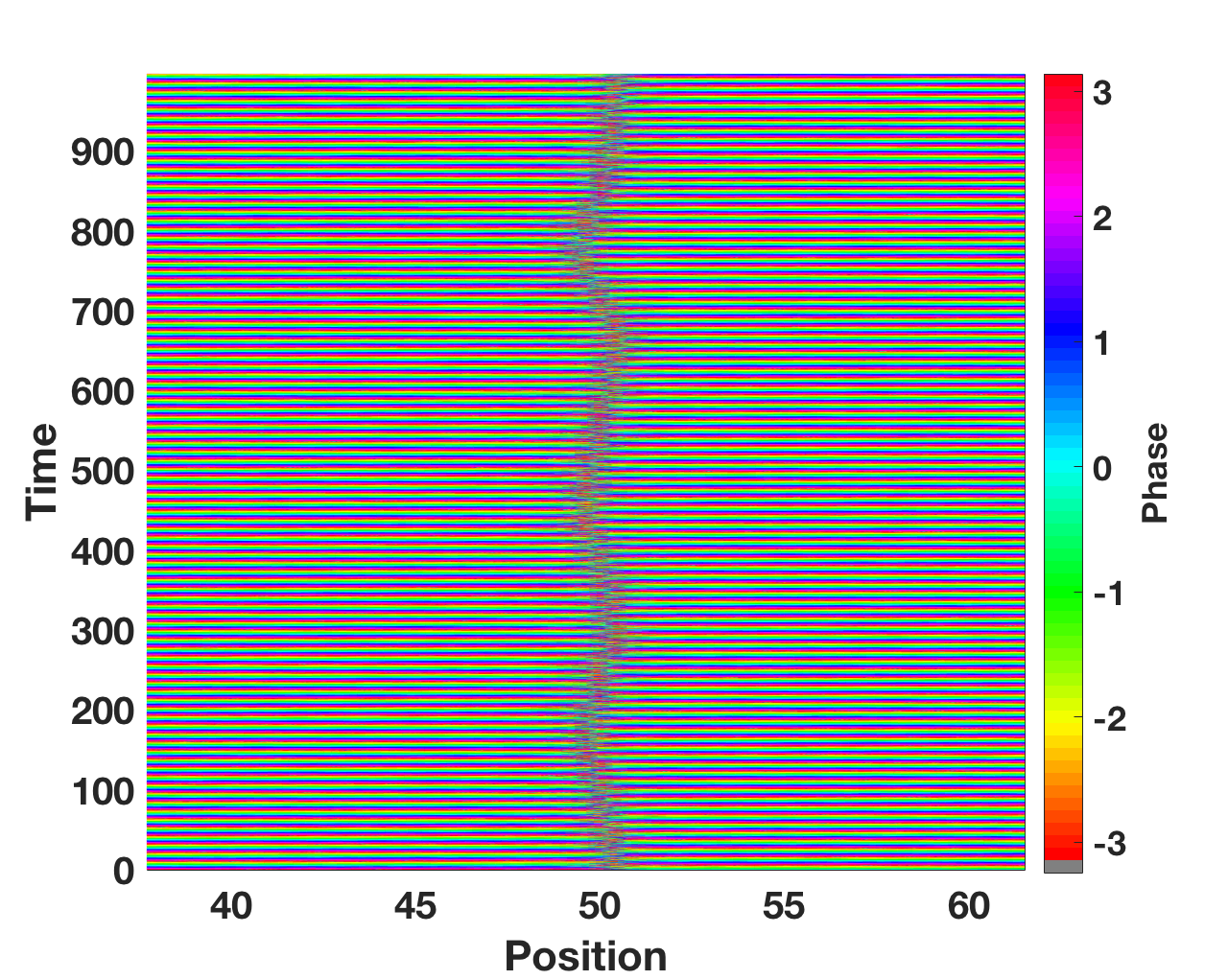}
\end{subfigure}
\caption[\emph{Oscillation of dark component in a harmonic potential with white noise added}.]{\emph{Oscillation of dark component in a harmonic potential with white noise added}. Shown are density (phase) in the upper (lower) panel for the dark component in a dark-bright soliton for the same parameters in Fig.~\ref{fig:FRHPRA:DS_den_no_noise}  with $5\%$ noise added to the initial wave function at $t=0$. The dark component oscillates with the same frequency in Fig.~\ref{fig:FRHPRA:DS_den_no_noise} but with a slightly smaller oscillation amplitude.}
\label{fig:FRHPRA:DS_den_with_noise}
\end{figure}

In this section, we address the question of experimental observability.  How stable are the dominant frequencies of dark-bright soliton motion in a harmonic trap?  To answer this questions, we add white noise to the system in the spatial Fourier transform of the initial condition at the 5\% level, then reverse Fourier transform to obtain a noisy initial state. Propagating this noisy initial state, we plot two cases for the same harmonic potential trap frequency, $\Omega/2\pi = 0.0064$ in Fig.~\ref{fig:FRHPRA:BS_den_no_noise}, ~\ref{fig:FRHPRA:DS_den_no_noise}, ~\ref{fig:FRHPRA:BS_den_with_noise} and ~\ref{fig:FRHPRA:DS_den_with_noise}. The free parameters are again taken to match our test case used throughout this paper, although we also examined other cases to find similar features. In Fig.~\ref{fig:FRHPRA:BS_den_no_noise} and ~\ref{fig:FRHPRA:DS_den_no_noise} we plot the density of the bright component and the dark component, respectively. The dark-bright soliton oscillates with $\omega_{\mathrm{DB}}/ 2 \pi=0.0039$ and the internal oscillation in this case is $\omega_{\mathrm{internal}}=0.032$. The noisy case is found to oscillate with the same frequency but with a slightly reduced oscillation amplitudeas can be seen in Fig.~\ref{fig:FRHPRA:BS_den_with_noise} for the bright component and in Fig.~\ref{fig:FRHPRA:DS_den_with_noise} for the dark component.  Thus we expect our predictions to be experimentally observable.
%
\section{Conclusions}
  \label{sec:FRHPRA:Conclusions}
We obtained a system of equation of motions for a dark-bright soliton in a harmonic potential. We used a variational method with a hyperbolic tangent for the dark component and a hyperbolic secant for the bright component. The harmonic potential modifies the background of the dark component according to the well-known Thomas-Fermi background approximation. A perturbation method was needed to include the effect of the harmonic potential, which amounts to restricting our analytical treatment cigar-shaped traps, also common in experiments.
%
%

The decoupling of relative and center of mass degrees of freedom for the harmonic case occurs for the classical two-body problem as well as its quantum extension, including to more than two particles, with relative coordinates appropriately generalized.  It is not immediately obvious this decoupling should also occur for a two-body bound state of two emergent features, a bright and a dark soliton.  For example, spontaneous symmetry breaking often causes such emergent properties to not respect underlying symmetries.  In previous work, we showed that for a uniform system the decoupling in fact does hold~\cite{majed2017}.  For a weak trap, this property nearly holds, but as the trapping strength is increased, internal oscillations and external motion are strongly coupled.  The effective potential, consisting of a sum between the potential and the mean field, may well be responsible for this effect, as found for example in non-exponential tunneling decay out of quasibound states in the scalar case~\cite{PhysRevA.96.063601,PhysRevLett.118.060402,0953-4075-38-17-012}.  As the trap is tightened the edges of the condensate are deformed by approach of the dark-bright soliton during its oscillations.  Because we treat a purely repulsive condensate in both components, the effective potential is larger than the bare potential, leading to a higher effective trapping frequency.  Moreover, the edges of the trap now impinge on the dark-bright soliton internal oscillations, shortening the internal oscillation time and therefore leading to a higher frequency.  The result is a coupling between center of mass motion deforming the effective potential, and internal oscillations being sped up by the deformation.

Future work could be the study of the internal oscillation of the two-component dark-bright soliton in a harmonic potential with an impurity at the center to look at the damping of a dark-bright soliton under periodic interaction with an impurity.  Other works have investigated the interaction of a dark-bright soliton in a harmonic potential with an impurity, but they did not take into account internal modes. Thus we suggest adding one more degree of freedom, namely, a relative coordinate for the position of the dark and bright solitons, which as we have shown is vital to understand and predict harmonic motion.


\appendix
\section{Matrix elements}
\label{appendix:Matrix elements}
The matrix elements in Eq.~\eqref{eq:matrix1} are,
\begin{align}
	\label{eq:matrix_elements_1}
	A_{11} &= i \omega  , \\ \nonumber
	A_{12} &= \frac{8 g u^2_{0}}{15 g_{1} \mathrm{w}^2_{\mathrm{fp}}}- \frac{g \pi^2 \Omega^2}{45 g_{1} }  , \\ \nonumber
	A_{13} &= -\frac{8 g u^2_{0}}{15 g_{1} \mathrm{w}^2_{\mathrm{fp}}} - \Omega^2 + \frac{2 g \Omega^2}{3 g_{1}} + \frac{ g \pi^2 \Omega^2}{45 g_{1}} , \\ \nonumber
	A_{22} &=  \frac{4 g u^2_{0}}{15 g_{2} \mathrm{w}^2_{\mathrm{fp}}}-\frac{\Omega^2 (2+u^2_{0}  \mathrm{w}^2_{\mathrm{fp}} )}{6 u^2_{0} \mathrm{w}_{\mathrm{fp}} }, \\ \nonumber
	A_{23} &=  -\frac{4 g u^2_{0}}{15 g_{2} \mathrm{w}^2_{\mathrm{fp}}}, \\ \nonumber
	A_{24} &=  i \omega, \\ \nonumber
	A_{32} &=  \frac{4 i u^2_{0} \omega}{g_{1}}, \\ \nonumber
	A_{34} &=  -\frac{\pi^2 u^2_{0} \mathrm{w}^3_{\mathrm{fp}} \Omega^2}{6 g_{1}}, \\ \nonumber
	A_{45} &=  \frac{16 g u^4_{0} \mathrm{w}_{\mathrm{fp}}}{3g_{1} g_{2}}, \\ \nonumber
	A_{46} &=  -\frac{8 u^4_{0}}{3 g_{1}} + \frac{8 g u^4_{0}}{3 g_{1} g_{2}} + \frac{4  u^2_{0}}{3 g_{1} \mathrm{w}^2_{\mathrm{fp}}}
	 \\ \nonumber & - \Omega^2(\frac{6+(12+\pi^2) u^2_{0} \mathrm{w}^2_{\mathrm{fp}} }{6 g_{1}}) , \\ \nonumber
	A_{47} &=  -\frac{4 u^2_{0}}{3 g_{1} \mathrm{w}_{\mathrm{fp}}} - \frac{8  u^4_{0} \mathrm{w}_{\mathrm{fp}}}{g_{1}} + \frac{8 g  u^4_{0} \mathrm{w}_{\mathrm{fp}}}{3 g_{1} g_{1} }
	 \\ \nonumber &- \Omega^2(\frac{6+(12+\pi^2) u^2_{0} \mathrm{w}^3_{\mathrm{fp}} }{6 g_{1}}), \\ \nonumber
	A_{55} &=  2 i  \mathrm{w}_{\mathrm{fp}} \omega, \\ \nonumber
	A_{56} &=  i \omega, \\ \nonumber
	A_{61} &=  \mathrm{w}_{\mathrm{fp}}, \\ \nonumber
	A_{63} &=  i \mathrm{w}_{\mathrm{fp}} \omega, \\ \nonumber
	A_{65} &=  \mathrm{w}_{\mathrm{fp}} \phi_{1fp} , \\ \nonumber
	A_{66} &=  \phi_{1fp} , \\ \nonumber
	A_{75} &=  -\frac{4(g_{1}-g_{2})u^2_{0}}{3 g_{1} g_{2} \mathrm{w}_{\mathrm{fp}}}  -\frac{4(3g_{1}+g_{2})u^4_{0} \mathrm{w}_{\mathrm{fp}} }{3 g_{1} g_{2} }  \\ \nonumber &
	+ \Omega^2 \left(\frac{2(\pi^2 - 6) \mathrm{w}_{\mathrm{fp}} }{9 g_{1}} \right. \\ \nonumber & \left. + \frac{u^2_{0} \mathrm{w}^3_{\mathrm{fp}} }{18 g_{1} g_{2}}
	 \left[-6(g+2g_{2})  +\pi^2(-5g+6g_{1}+3g_{2})\right] \right), \\ \nonumber
	A_{76} &=  -\frac{4(g_{1}-g_{2})u^4_{0}}{3 g_{1} g_{2} }  +\frac{4(g_{1}+g_{2})u^2_{0}  }{3 g_{1} g_{2} \mathrm{w}^2_{\mathrm{fp}}}   \\ \nonumber &
	+\Omega^2 \left(\frac{12-2\pi^2+27 u^2_{0} \mathrm{w}^2_{\mathrm{fp}} }{9 g_{1}} +   \right. \\ \nonumber & \left.  \frac{(2 g_{1}-g_{2})\pi^2 u^2_{0} \mathrm{w}^2_{\mathrm{fp}}}{2 g_{1} g_{2}} -\frac{g(6+5 \pi^2) u^2_{0} \mathrm{w}^2_{\mathrm{fp}} }{6 g_{1} g_{2}} \right)   , \\ \nonumber	
\end{align}	
\begin{align}
	\label{eq:matrix_elements_2}
	A_{77} &=  \frac{8 u^2_{0}}{3 g_{1} \mathrm{w}_{\mathrm{fp}}} (-1+2 u^2_{0} \mathrm{w}^2_{\mathrm{fp}}) + \\ \nonumber & \Omega^2 \left(-\frac{2(-6+\pi^2)\mathrm{w}_{\mathrm{fp}} (2+3 u^2_{0} \mathrm{w}^2_{\mathrm{fp}})}{9 g_{1}} \right. \\& \nonumber \left.
	+\frac{g(-6+\pi^2) u^2_{0} \mathrm{w}^3_{\mathrm{fp}}}{9 g_{1} g_{2}}\right), \\ \nonumber
	A_{85} &= -\frac{2 u^2_{0}}{3 g_{1} g_{2} \mathrm{w}_{\mathrm{fp}}} (g_{1}+2 g u^2_{0} \mathrm{w}^2_{\mathrm{fp}} +12 g_{1} u^2_{0} \mathrm{w}^2_{\mathrm{fp}}) + \\ \nonumber & \Omega^2 (\frac{g(12+\pi^2) u^2_{0} \mathrm{w}^3_{\mathrm{fp}}}{18 g_{1} g_{2}}-\frac{\pi^2 u^2_{0} \mathrm{w}^3_{\mathrm{fp}}}{6 g_{2}}) , \\ \nonumber
	A_{86} &=  \frac{1}{2 g_{2} g_{2} \mathrm{w}^2_{\mathrm{fp}}} (2 g_{1} u^2_{0} - 4 g u^4_{0} \mathrm{w}^2_{\mathrm{fp}}-8 g_{1} u^4_{0} \mathrm{w}_{\mathrm{fp}}) + \\ \nonumber & \Omega^2 (\frac{g u^2_{0} \mathrm{w}^2_{\mathrm{fp}} (12+\pi^2)}{6 g_{1} g_{2}} -\frac{\pi^2 u^2_{0} \mathrm{w}^2_{\mathrm{fp}}}{2 g_{2}}), \\ \nonumber
	A_{87} &=  \frac{16 g u^4_{0} \mathrm{w}_{\mathrm{fp}}}{3 g_{1} g_{2}} - \Omega^2 (\frac{2 g u^2_{0} \mathrm{w}^3_{\mathrm{fp}}}{9 g_{1} g_{2}}), \\ \nonumber
	A_{88} &=  -\frac{4 i u^2_{0} \mathrm{w}_{\mathrm{fp}} \omega}{g_{2}}. \\ \nonumber
\end{align}	
The frequency coefficients are,
\begin{align}
	\label{eq:constants_frequency}
	&\alpha_{1} = \frac{1}{27 g^3_{1} g^3_{2} } \left[768 u^8_{0} (3g_{2} (g_{2}-g_{1}) \right. \\& \nonumber \left. + u^2_{0} \mathrm{w}^2_{\mathrm{fp}} (2g(3g_{1}+g_{2})-g_{2} (19g_{1}+25g_{2})) \right. \\& \nonumber \left.+ 2 (-g_{2}(3g_{1}+g_{2})+g(g_{1}+3g_{2})) u^4_{0} \mathrm{w}^4_{\mathrm{fp}} \right. \\& \nonumber \left.
	- 32 u^6_{0}  \mathrm{w}^2_{\mathrm{fp}} \{2 g_{2} (27 g_{1} + g_{2} (129-14 \pi^2 ) ) \right. \\& \nonumber \left.
	+ g_{2} (-198 g +126 g_{1} +600 g_{2} + (3 g+39 g_{1}) \right. \\& \nonumber \left.
	-70 g_{2} \pi^2)  u^2_{0} \mathrm{w}^2_{\mathrm{fp}}  + 12 (5 g^2+32 g g_{2} +3 g_{2} (g_{1}+8g_{2})) \right. \\& \nonumber \left.
	+  u^4_{0} \mathrm{w}^4_{\mathrm{fp}} (50 g^2 +3(61 g_{1} -31 g_{2} )g_{2}  \right. \\& \nonumber \left.
	- 4 g (15 g_{1}+29 g_{2}) ) \} \Omega^2 \right. \\& \nonumber \left.
	-4 g_{2} u^4_{0} \mathrm{w}^4_{\mathrm{fp}} (6+(12+\pi^2) u^2_{0} \mathrm{w}^2_{\mathrm{fp}}) (4 g_{2}(-6+\pi^2)) \right. \\& \nonumber \left.
	 + (-6 (g+3g_{2}) + (-29 g +30 g_{1} +3 g_{2})\pi^2 u^2_{0}) \mathrm{w}^2_{\mathrm{fp}}\Omega^4 )\right] , \\ \nonumber
	&\alpha_{2} = \frac{u^2_{0}}{4860 g^4_{1} g^4_{2}  \mathrm{w}^2_{\mathrm{fp}}}  \{ 384 g g_{2} u^4_{0} + 8 u^2_{0}  \mathrm{w}^2_{\mathrm{fp}} (90 g_{1} g_{2} \\& \nonumber
	-2 g g_{2} (30+\pi^2) + g g_{1} \pi^2 u^2_{0}  \mathrm{w}^2_{\mathrm{fp}}) \Omega^2  \\& \nonumber
	-5 g_{1} g_{2} \pi^2 \mathrm{w}^4_{\mathrm{fp}} (2+u^2_{0} \mathrm{w}^2_{\mathrm{fp}}) \Omega^4 \\& \nonumber
	\left[ 192 u^4_{0} (3(g_{1}-g_{2})g_{2}+(-2g(3g_{1}+g_{2}) \right. \\& \nonumber \left.
	+g_{2}(19g_{1}+25g_{2})) u^2_{0}  \mathrm{w}^2_{\mathrm{fp}} 2(g_{2}(3g_{1}+g_{2})-g (g_{1} \right. \\& \nonumber  \left.
	+3 g_{2}))u^4_{0}  \mathrm{w}^4_{\mathrm{fp}})\right] \\& \nonumber
	+ 8 u^2_{0}  \mathrm{w}^2_{\mathrm{fp}} ( 2 g_{2} ( 27 g_{1}  \\& \nonumber
	+ g_{2} ( 129- 14 \pi^2 )) + g_{2} ( -198 g + 126 g_{1}  \\& \nonumber
	+ 600 g_{2} + (3 g + 39 g_{1} -70 g_{2} ) \pi^2) u^2_{0} \mathrm{w}^2_{\mathrm{fp}} + (12 (5 g^2 \\& \nonumber
	 -32 g g^2 + 3 g_{2} (g_{1}+8g_{2}) + (50 g^2 +3 (61 g_{1} -31 g_{2} ) g_{2} \\& \nonumber
	 - 4 g (15 g_{1}+29 g_{2}) ) \pi^2 ) u^4_{0} \mathrm{w}^4_{\mathrm{fp}}) \Omega^2 \\& \nonumber
	 +g_{2} \mathrm{w}^4_{\mathrm{fp}} (^ + ( 12 + \pi^2 ) u^2_{0} \mathrm{w}^2_{\mathrm{fp}} ) ( 4 g_{2} (-6+\pi^2) +(-6(g+3g_{2}) \\& \nonumber
	 +(-29g+30g_{1}+3g_{2})\pi^2) u^4_{0}  \mathrm{w}^4_{\mathrm{fp}} ) \Omega^4)
	\}
	, \\ \nonumber
	&\alpha_{3} = \frac{\pi^2 u^2_{0} \Omega^4}{43740 g^4_{1} g^4_{2}} ( 24 g u^2_{0} (2g_{2}+(2g-3g_{1}+g_{2}) u^2_{0} \mathrm{w}^2_{\mathrm{fp}})\\& \nonumber
	-g_{2}(-45 g_{1} + g(30+\pi^2)) \mathrm{w}^2_{\mathrm{fp}} (2+u^2_{0} \mathrm{w}^2_{\mathrm{fp}}) \Omega^2) \\& \nonumber
	(192 u^4_{0} (3(g_{1} - g_{2}) g_{2} +(-2 g(3g_{1}+g_{2}) \\& \nonumber
	+g_{2}(19g_{1} + 25 g_{2})) u^2_{0}  \mathrm{w}^2_{\mathrm{fp}} \\& \nonumber
	 + 2 (g_{2} (3g_{1} +g_{2}) - g (g_{1} + 3 g_{2}) ) u^4_{0} \mathrm{w}^4_{\mathrm{fp}} )\\& \nonumber
	+ 8 u^2_{0} \mathrm{w}^2_{\mathrm{fp}} (2 g_{2} (27g_{1} +g_{2} (129-14\pi^2)) \\& \nonumber
	+ g_{2} (-198 g + 126 g_{1} \\& \nonumber
	+600 g_{2} + (3g+39g_{1}-70 g_{2})\pi^2)  u^4_{0} \mathrm{w}^4_{\mathrm{fp}} \\& \nonumber
	+(12 ( 5 g^2 -32 g g_{2} +3 g_{2} (g_{1}+8g_{2}))+(50 g^2 \\& \nonumber
	+3 (61 g_{1} -31 g_{2} ) g_{2} - 4 g (15 g_{1} +29 g_{2}) ) \pi^2 ) u^4_{0} \mathrm{w}^4_{\mathrm{fp}} ) \Omega ^2 \\& \nonumber
	g_{2} \mathrm{w}^4_{\mathrm{fp}} (6+(12+\pi^2)u^2_{0} \mathrm{w}^2_{\mathrm{fp}}) (4 g_{2} \\& \nonumber
	 (-6+\pi^2) + (-6(g+3g_{2})\\& \nonumber
	 +(-29g+30g1+3g_{2})\pi^2 ) u^2_{0} \mathrm{w}^2_{\mathrm{fp}} ) \Omega^4).
	\\ \nonumber
\end{align}	

\FloatBarrier
\newpage
\bibliography{library}

\begin{thebibliography}{27}%
\makeatletter
\providecommand \@ifxundefined [1]{%
 \@ifx{#1\undefined}
}%
\providecommand \@ifnum [1]{%
 \ifnum #1\expandafter \@firstoftwo
 \else \expandafter \@secondoftwo
 \fi
}%
\providecommand \@ifx [1]{%
 \ifx #1\expandafter \@firstoftwo
 \else \expandafter \@secondoftwo
 \fi
}%
\providecommand \natexlab [1]{#1}%
\providecommand \enquote  [1]{``#1''}%
\providecommand \bibnamefont  [1]{#1}%
\providecommand \bibfnamefont [1]{#1}%
\providecommand \citenamefont [1]{#1}%
\providecommand \href@noop [0]{\@secondoftwo}%
\providecommand \href [0]{\begingroup \@sanitize@url \@href}%
\providecommand \@href[1]{\@@startlink{#1}\@@href}%
\providecommand \@@href[1]{\endgroup#1\@@endlink}%
\providecommand \@sanitize@url [0]{\catcode `\\12\catcode `\$12\catcode
  `\&12\catcode `\#12\catcode `\^12\catcode `\_12\catcode `\%12\relax}%
\providecommand \@@startlink[1]{}%
\providecommand \@@endlink[0]{}%
\providecommand \url  [0]{\begingroup\@sanitize@url \@url }%
\providecommand \@url [1]{\endgroup\@href {#1}{\urlprefix }}%
\providecommand \urlprefix  [0]{URL }%
\providecommand \Eprint [0]{\href }%
\providecommand \doibase [0]{http://dx.doi.org/}%
\providecommand \selectlanguage [0]{\@gobble}%
\providecommand \bibinfo  [0]{\@secondoftwo}%
\providecommand \bibfield  [0]{\@secondoftwo}%
\providecommand \translation [1]{[#1]}%
\providecommand \BibitemOpen [0]{}%
\providecommand \bibitemStop [0]{}%
\providecommand \bibitemNoStop [0]{.\EOS\space}%
\providecommand \EOS [0]{\spacefactor3000\relax}%
\providecommand \BibitemShut  [1]{\csname bibitem#1\endcsname}%
\let\auto@bib@innerbib\@empty
\bibitem [{\citenamefont {Dauxois}\ and\ \citenamefont
  {Peyrard}(2006)}]{dauxois2006physics}%
  \BibitemOpen
  \bibfield  {author} {\bibinfo {author} {\bibfnamefont {T.}~\bibnamefont
  {Dauxois}}\ and\ \bibinfo {author} {\bibfnamefont {M.}~\bibnamefont
  {Peyrard}},\ }\href {https://books.google.com/books?id=YKe1UZc_Qo8C} {\emph
  {\bibinfo {title} {{Physics of Solitons}}}}\ (\bibinfo  {publisher}
  {Cambridge University Press},\ \bibinfo {year} {2006})\BibitemShut {NoStop}%
\bibitem [{\citenamefont {Kevrekidis}\ \emph {et~al.}(2008)\citenamefont
  {Kevrekidis}, \citenamefont {Frantzeskakis}, \citenamefont
  {Carretero-Gonz{\'{a}}lez}, \citenamefont {Parker}, \citenamefont {Jackson},
  \citenamefont {Martin},\ and\ \citenamefont {Adams}}]{Kevrekidis2008}%
  \BibitemOpen
  \bibfield  {author} {\bibinfo {author} {\bibfnamefont {P.~G.}\ \bibnamefont
  {Kevrekidis}}, \bibinfo {author} {\bibfnamefont {D.~J.}\ \bibnamefont
  {Frantzeskakis}}, \bibinfo {author} {\bibfnamefont {R.}~\bibnamefont
  {Carretero-Gonz{\'{a}}lez}}, \bibinfo {author} {\bibfnamefont {N.~G.}\
  \bibnamefont {Parker}}, \bibinfo {author} {\bibfnamefont {B.}~\bibnamefont
  {Jackson}}, \bibinfo {author} {\bibfnamefont {A.~M.}\ \bibnamefont {Martin}},
  \ and\ \bibinfo {author} {\bibfnamefont {C.~S.}\ \bibnamefont {Adams}},\
  }\href {\doibase 10.1007/978-3-540-73591-5} {\emph {\bibinfo {title}
  {{Emergent Nonlinear Phenomena in Bose-Einstein Condensates}}}}\ (\bibinfo
  {publisher} {Springer Series},\ \bibinfo {year} {2008})\BibitemShut {NoStop}%
\bibitem [{\citenamefont {Kevrekidis}\ \emph {et~al.}(2015)\citenamefont
  {Kevrekidis}, \citenamefont {Frantzeskakis},\ and\ \citenamefont
  {Carretero-Gonz{\'{a}}lez}}]{Kevrekidis2015}%
  \BibitemOpen
  \bibfield  {author} {\bibinfo {author} {\bibfnamefont {P.~G.}\ \bibnamefont
  {Kevrekidis}}, \bibinfo {author} {\bibfnamefont {D.~J.}\ \bibnamefont
  {Frantzeskakis}}, \ and\ \bibinfo {author} {\bibfnamefont {R.}~\bibnamefont
  {Carretero-Gonz{\'{a}}lez}},\ }\href
  {https://books.google.com/books?id=1BVoCgAAQBAJ} {\emph {\bibinfo {title}
  {{The Defocusing Nonlinear Schrodinger Equation: From Dark Solitons to
  Vortices and Vortex Rings}}}}\ (\bibinfo  {publisher} {Society for Industrial
  and Applied Mathematics},\ \bibinfo {year} {2015})\BibitemShut {NoStop}%
\bibitem [{\citenamefont {Pethick}\ and\ \citenamefont
  {Smith}(2008)}]{Pethick2008}%
  \BibitemOpen
  \bibfield  {author} {\bibinfo {author} {\bibfnamefont {C.~J.}\ \bibnamefont
  {Pethick}}\ and\ \bibinfo {author} {\bibfnamefont {H.}~\bibnamefont
  {Smith}},\ }\href {\doibase DOI: 10.1017/CBO9780511802850} {\emph {\bibinfo
  {title} {{Bose–Einstein Condensation in Dilute Gases}}}},\ \bibinfo
  {edition} {2nd}\ ed.\ (\bibinfo  {publisher} {Cambridge University Press},\
  \bibinfo {address} {Cambridge},\ \bibinfo {year} {2008})\BibitemShut
  {NoStop}%
\bibitem [{\citenamefont {Cuevas}\ \emph {et~al.}(2012)\citenamefont {Cuevas},
  \citenamefont {Chang}, \citenamefont {Hamner}, \citenamefont {Hoefer},
  \citenamefont {Kevrekidis}, \citenamefont {Engels}, \citenamefont
  {Achilleos}, \citenamefont {Frantzeskakis},\ and\ \citenamefont
  {J}}]{Cuevas2012}%
  \BibitemOpen
  \bibfield  {author} {\bibinfo {author} {\bibfnamefont {D.~Y.}\ \bibnamefont
  {Cuevas}}, \bibinfo {author} {\bibfnamefont {J.~J.}\ \bibnamefont {Chang}},
  \bibinfo {author} {\bibfnamefont {C.}~\bibnamefont {Hamner}}, \bibinfo
  {author} {\bibfnamefont {M.}~\bibnamefont {Hoefer}}, \bibinfo {author}
  {\bibfnamefont {P.~G.}\ \bibnamefont {Kevrekidis}}, \bibinfo {author}
  {\bibfnamefont {P.}~\bibnamefont {Engels}}, \bibinfo {author} {\bibfnamefont
  {V.}~\bibnamefont {Achilleos}}, \bibinfo {author} {\bibfnamefont {D.~J.}\
  \bibnamefont {Frantzeskakis}}, \ and\ \bibinfo {author} {\bibnamefont {J}},\
  }\href {http://stacks.iop.org/0953-4075/45/i=11/a=115301} {\bibfield
  {journal} {\bibinfo  {journal} {Journal of Physics B: Atomic, Molecular and
  Optical Physics}\ }\textbf {\bibinfo {volume} {45}},\ \bibinfo {pages}
  {115301} (\bibinfo {year} {2012})}\BibitemShut {NoStop}%
\bibitem [{\citenamefont {Liu}\ \emph {et~al.}(2012)\citenamefont {Liu},
  \citenamefont {Lu},\ and\ \citenamefont {Liu}}]{Liu2012a}%
  \BibitemOpen
  \bibfield  {author} {\bibinfo {author} {\bibfnamefont {C.-F.}\ \bibnamefont
  {Liu}}, \bibinfo {author} {\bibfnamefont {M.}~\bibnamefont {Lu}}, \ and\
  \bibinfo {author} {\bibfnamefont {W.-Q.}\ \bibnamefont {Liu}},\ }\href
  {http://www.sciencedirect.com/science/article/pii/S0375960111013272}
  {\bibfield  {journal} {\bibinfo  {journal} {Physics Letters A}\ }\textbf
  {\bibinfo {volume} {376}},\ \bibinfo {pages} {188} (\bibinfo {year}
  {2012})}\BibitemShut {NoStop}%
\bibitem [{\citenamefont {Hoefer}\ \emph {et~al.}(2011)\citenamefont {Hoefer},
  \citenamefont {Chang}, \citenamefont {Hamner},\ and\ \citenamefont
  {Engels}}]{Hoefer2011b}%
  \BibitemOpen
  \bibfield  {author} {\bibinfo {author} {\bibfnamefont {M.~A.}\ \bibnamefont
  {Hoefer}}, \bibinfo {author} {\bibfnamefont {J.~J.}\ \bibnamefont {Chang}},
  \bibinfo {author} {\bibfnamefont {C.}~\bibnamefont {Hamner}}, \ and\ \bibinfo
  {author} {\bibfnamefont {P.}~\bibnamefont {Engels}},\ }\href
  {http://link.aps.org/doi/10.1103/PhysRevA.84.041605} {\bibfield  {journal}
  {\bibinfo  {journal} {Physical Review A}\ }\textbf {\bibinfo {volume} {84}},\
  \bibinfo {pages} {41605} (\bibinfo {year} {2011})}\BibitemShut {NoStop}%
\bibitem [{\citenamefont {Xun-Xu}\ \emph {et~al.}(2011)\citenamefont {Xun-Xu},
  \citenamefont {Pei}, \citenamefont {Wan-Quan},\ and\ \citenamefont
  {Liu}}]{Xun-Xu2011}%
  \BibitemOpen
  \bibfield  {author} {\bibinfo {author} {\bibfnamefont {Z.~X.-F.}\
  \bibnamefont {Xun-Xu}}, \bibinfo {author} {\bibfnamefont {Z.}~\bibnamefont
  {Pei}}, \bibinfo {author} {\bibfnamefont {H.}~\bibnamefont {Wan-Quan}}, \
  and\ \bibinfo {author} {\bibnamefont {Liu}},\ }\href
  {http://stacks.iop.org/1674-1056/20/i=2/a=020307} {\bibfield  {journal}
  {\bibinfo  {journal} {Chinese Physics B}\ }\textbf {\bibinfo {volume} {20}},\
  \bibinfo {pages} {20307} (\bibinfo {year} {2011})}\BibitemShut {NoStop}%
\bibitem [{\citenamefont {Becker}\ \emph {et~al.}(2008)\citenamefont {Becker},
  \citenamefont {Stellmer}, \citenamefont {Soltan-Panahi}, \citenamefont
  {Dorscher}, \citenamefont {Baumert}, \citenamefont {Richter}, \citenamefont
  {Kronjager}, \citenamefont {Bongs},\ and\ \citenamefont
  {Sengstock}}]{Becker2008b}%
  \BibitemOpen
  \bibfield  {author} {\bibinfo {author} {\bibfnamefont {C.}~\bibnamefont
  {Becker}}, \bibinfo {author} {\bibfnamefont {S.}~\bibnamefont {Stellmer}},
  \bibinfo {author} {\bibfnamefont {P.}~\bibnamefont {Soltan-Panahi}}, \bibinfo
  {author} {\bibfnamefont {S.}~\bibnamefont {Dorscher}}, \bibinfo {author}
  {\bibfnamefont {M.}~\bibnamefont {Baumert}}, \bibinfo {author} {\bibfnamefont
  {E.-M.}\ \bibnamefont {Richter}}, \bibinfo {author} {\bibfnamefont
  {J.}~\bibnamefont {Kronjager}}, \bibinfo {author} {\bibfnamefont
  {K.}~\bibnamefont {Bongs}}, \ and\ \bibinfo {author} {\bibfnamefont
  {K.}~\bibnamefont {Sengstock}},\ }\href {http://dx.doi.org/10.1038/nphys962}
  {\bibfield  {journal} {\bibinfo  {journal} {Nat Phys}\ }\textbf {\bibinfo
  {volume} {4}},\ \bibinfo {pages} {496} (\bibinfo {year} {2008})}\BibitemShut
  {NoStop}%
\bibitem [{\citenamefont {Busch}\ and\ \citenamefont
  {Anglin}(2001)}]{Busch2001}%
  \BibitemOpen
  \bibfield  {author} {\bibinfo {author} {\bibfnamefont {T.}~\bibnamefont
  {Busch}}\ and\ \bibinfo {author} {\bibfnamefont {J.~R.}\ \bibnamefont
  {Anglin}},\ }\href {https://link.aps.org/doi/10.1103/PhysRevLett.87.010401}
  {\bibfield  {journal} {\bibinfo  {journal} {Physical Review Letters}\
  }\textbf {\bibinfo {volume} {87}},\ \bibinfo {pages} {10401} (\bibinfo {year}
  {2001})}\BibitemShut {NoStop}%
\bibitem [{\citenamefont {Frantzeskakis}\ \emph {et~al.}(2012)\citenamefont
  {Frantzeskakis}, \citenamefont {Yan}, \citenamefont {Kevrekidis},\ and\
  \citenamefont {J}}]{Frantzeskakis2012}%
  \BibitemOpen
  \bibfield  {author} {\bibinfo {author} {\bibfnamefont {V.~A.}\ \bibnamefont
  {Frantzeskakis}}, \bibinfo {author} {\bibfnamefont {D.}~\bibnamefont {Yan}},
  \bibinfo {author} {\bibfnamefont {P.~G.}\ \bibnamefont {Kevrekidis}}, \ and\
  \bibinfo {author} {\bibfnamefont {D.}~\bibnamefont {J}},\ }\href
  {http://stacks.iop.org/1367-2630/14/i=5/a=055006} {\bibfield  {journal}
  {\bibinfo  {journal} {New Journal of Physics}\ }\textbf {\bibinfo {volume}
  {14}},\ \bibinfo {pages} {55006} (\bibinfo {year} {2012})}\BibitemShut
  {NoStop}%
\bibitem [{\citenamefont {Hamner}\ \emph {et~al.}(2011)\citenamefont {Hamner},
  \citenamefont {Chang}, \citenamefont {Engels},\ and\ \citenamefont
  {Hoefer}}]{Hamner2011b}%
  \BibitemOpen
  \bibfield  {author} {\bibinfo {author} {\bibfnamefont {C.}~\bibnamefont
  {Hamner}}, \bibinfo {author} {\bibfnamefont {J.~J.}\ \bibnamefont {Chang}},
  \bibinfo {author} {\bibfnamefont {P.}~\bibnamefont {Engels}}, \ and\ \bibinfo
  {author} {\bibfnamefont {M.~A.}\ \bibnamefont {Hoefer}},\ }\href
  {http://link.aps.org/doi/10.1103/PhysRevLett.106.065302} {\bibfield
  {journal} {\bibinfo  {journal} {Physical Review Letters}\ }\textbf {\bibinfo
  {volume} {106}},\ \bibinfo {pages} {65302} (\bibinfo {year}
  {2011})}\BibitemShut {NoStop}%
\bibitem [{\citenamefont {Rajendran}\ \emph {et~al.}(2009)\citenamefont
  {Rajendran}, \citenamefont {Muruganandam},\ and\ \citenamefont
  {Lakshmanan}}]{Rajendran2009b}%
  \BibitemOpen
  \bibfield  {author} {\bibinfo {author} {\bibfnamefont {S.}~\bibnamefont
  {Rajendran}}, \bibinfo {author} {\bibfnamefont {P.}~\bibnamefont
  {Muruganandam}}, \ and\ \bibinfo {author} {\bibfnamefont {M.}~\bibnamefont
  {Lakshmanan}},\ }\href {\doibase 10.1088/0953-4075/42/14/145307} {\bibfield
  {journal} {\bibinfo  {journal} {Journal of Physics B: Atomic, Molecular and
  Optical Physics}\ }\textbf {\bibinfo {volume} {42}},\ \bibinfo {pages}
  {145307} (\bibinfo {year} {2009})}\BibitemShut {NoStop}%
\bibitem [{\citenamefont {Zhang}\ \emph {et~al.}(2009)\citenamefont {Zhang},
  \citenamefont {Hu}, \citenamefont {Liu},\ and\ \citenamefont
  {Liu}}]{Zhang2009b}%
  \BibitemOpen
  \bibfield  {author} {\bibinfo {author} {\bibfnamefont {X.-F.}\ \bibnamefont
  {Zhang}}, \bibinfo {author} {\bibfnamefont {X.-H.}\ \bibnamefont {Hu}},
  \bibinfo {author} {\bibfnamefont {X.-X.}\ \bibnamefont {Liu}}, \ and\
  \bibinfo {author} {\bibfnamefont {W.~M.}\ \bibnamefont {Liu}},\ }\href
  {http://link.aps.org/doi/10.1103/PhysRevA.79.033630} {\bibfield  {journal}
  {\bibinfo  {journal} {Physical Review A}\ }\textbf {\bibinfo {volume} {79}},\
  \bibinfo {pages} {33630} (\bibinfo {year} {2009})}\BibitemShut {NoStop}%
\bibitem [{\citenamefont {Strecker}\ \emph {et~al.}(2002)\citenamefont
  {Strecker}, \citenamefont {Partridge}, \citenamefont {Truscott},\ and\
  \citenamefont {Hulet}}]{Strecker2002b}%
  \BibitemOpen
  \bibfield  {author} {\bibinfo {author} {\bibfnamefont {K.~E.}\ \bibnamefont
  {Strecker}}, \bibinfo {author} {\bibfnamefont {G.~B.}\ \bibnamefont
  {Partridge}}, \bibinfo {author} {\bibfnamefont {A.~G.}\ \bibnamefont
  {Truscott}}, \ and\ \bibinfo {author} {\bibfnamefont {R.~G.}\ \bibnamefont
  {Hulet}},\ }\href {\doibase 10.1038/nature747} {\bibfield  {journal}
  {\bibinfo  {journal} {Nature}\ }\textbf {\bibinfo {volume} {417}},\ \bibinfo
  {pages} {150} (\bibinfo {year} {2002})}\BibitemShut {NoStop}%
\bibitem [{\citenamefont {Filatrella}\ \emph {et~al.}(2014)\citenamefont
  {Filatrella}, \citenamefont {Malomed},\ and\ \citenamefont
  {Salerno}}]{Filatrella2014}%
  \BibitemOpen
  \bibfield  {author} {\bibinfo {author} {\bibfnamefont {G.}~\bibnamefont
  {Filatrella}}, \bibinfo {author} {\bibfnamefont {B.~a.}\ \bibnamefont
  {Malomed}}, \ and\ \bibinfo {author} {\bibfnamefont {M.}~\bibnamefont
  {Salerno}},\ }\href {\doibase 10.1103/PhysRevA.90.043629} {\bibfield
  {journal} {\bibinfo  {journal} {Physical Review A}\ }\textbf {\bibinfo
  {volume} {90}},\ \bibinfo {pages} {043629} (\bibinfo {year}
  {2014})}\BibitemShut {NoStop}%
\bibitem [{\citenamefont {{Kh. ABDULLAEV}}\ \emph {et~al.}(2005)\citenamefont
  {{Kh. ABDULLAEV}}, \citenamefont {GAMMAL}, \citenamefont {KAMCHATNOV},\ and\
  \citenamefont {TOMIO}}]{doi:10.1142/S0217979205032279}%
  \BibitemOpen
  \bibfield  {author} {\bibinfo {author} {\bibfnamefont {F.}~\bibnamefont {{Kh.
  ABDULLAEV}}}, \bibinfo {author} {\bibfnamefont {A.}~\bibnamefont {GAMMAL}},
  \bibinfo {author} {\bibfnamefont {A.~M.}\ \bibnamefont {KAMCHATNOV}}, \ and\
  \bibinfo {author} {\bibfnamefont {L.}~\bibnamefont {TOMIO}},\ }\href
  {\doibase 10.1142/S0217979205032279} {\bibfield  {journal} {\bibinfo
  {journal} {International Journal of Modern Physics B}\ }\textbf {\bibinfo
  {volume} {19}},\ \bibinfo {pages} {3415} (\bibinfo {year}
  {2005})}\BibitemShut {NoStop}%
\bibitem [{\citenamefont {P{\'{e}}rez-Garc{\'{i}}a}\ and\ \citenamefont
  {Beitia}(2005)}]{Perez-Garcia2005}%
  \BibitemOpen
  \bibfield  {author} {\bibinfo {author} {\bibfnamefont {V.}~\bibnamefont
  {P{\'{e}}rez-Garc{\'{i}}a}}\ and\ \bibinfo {author} {\bibfnamefont
  {J.}~\bibnamefont {Beitia}},\ }\href {\doibase 10.1103/PhysRevA.72.033620}
  {\bibfield  {journal} {\bibinfo  {journal} {Physical Review A}\ }\textbf
  {\bibinfo {volume} {72}},\ \bibinfo {pages} {1} (\bibinfo {year}
  {2005})}\BibitemShut {NoStop}%
\bibitem [{\citenamefont {Morera}\ \emph {et~al.}(2018)\citenamefont {Morera},
  \citenamefont {Mateo}, \citenamefont {Polls},\ and\ \citenamefont
  {Juli\'a-D\'{\i}az}}]{PhysRevA.97.043621}%
  \BibitemOpen
  \bibfield  {author} {\bibinfo {author} {\bibfnamefont {I.}~\bibnamefont
  {Morera}}, \bibinfo {author} {\bibfnamefont {A.~M.~n.}\ \bibnamefont
  {Mateo}}, \bibinfo {author} {\bibfnamefont {A.}~\bibnamefont {Polls}}, \ and\
  \bibinfo {author} {\bibfnamefont {B.}~\bibnamefont {Juli\'a-D\'{\i}az}},\
  }\href {\doibase 10.1103/PhysRevA.97.043621} {\bibfield  {journal} {\bibinfo
  {journal} {Phys. Rev. A}\ }\textbf {\bibinfo {volume} {97}},\ \bibinfo
  {pages} {043621} (\bibinfo {year} {2018})}\BibitemShut {NoStop}%
\bibitem [{\citenamefont {Achilleos}\ \emph {et~al.}(2011)\citenamefont
  {Achilleos}, \citenamefont {Kevrekidis}, \citenamefont {Rothos},\ and\
  \citenamefont {Frantzeskakis}}]{PhysRevA.84.053626}%
  \BibitemOpen
  \bibfield  {author} {\bibinfo {author} {\bibfnamefont {V.}~\bibnamefont
  {Achilleos}}, \bibinfo {author} {\bibfnamefont {P.~G.}\ \bibnamefont
  {Kevrekidis}}, \bibinfo {author} {\bibfnamefont {V.~M.}\ \bibnamefont
  {Rothos}}, \ and\ \bibinfo {author} {\bibfnamefont {D.~J.}\ \bibnamefont
  {Frantzeskakis}},\ }\href {\doibase 10.1103/PhysRevA.84.053626} {\bibfield
  {journal} {\bibinfo  {journal} {Phys. Rev. A}\ }\textbf {\bibinfo {volume}
  {84}},\ \bibinfo {pages} {53626} (\bibinfo {year} {2011})}\BibitemShut
  {NoStop}%
\bibitem [{\citenamefont {Alotaibi}\ and\ \citenamefont
  {Carr}(2017)}]{majed2017}%
  \BibitemOpen
  \bibfield  {author} {\bibinfo {author} {\bibfnamefont {M.~O.~D.}\
  \bibnamefont {Alotaibi}}\ and\ \bibinfo {author} {\bibfnamefont {L.~D.}\
  \bibnamefont {Carr}},\ }\href {\doibase 10.1103/PhysRevA.96.013601}
  {\bibfield  {journal} {\bibinfo  {journal} {Phys. Rev. A}\ }\textbf {\bibinfo
  {volume} {96}},\ \bibinfo {pages} {13601} (\bibinfo {year}
  {2017})}\BibitemShut {NoStop}%
\bibitem [{\citenamefont {Kivshar}\ and\ \citenamefont
  {Kr{\'{o}}likowski}(1995)}]{Kivshar1995}%
  \BibitemOpen
  \bibfield  {author} {\bibinfo {author} {\bibfnamefont {Y.}~\bibnamefont
  {Kivshar}}\ and\ \bibinfo {author} {\bibfnamefont {W.}~\bibnamefont
  {Kr{\'{o}}likowski}},\ }\href
  {http://www.sciencedirect.com/science/article/pii/003040189400644A}
  {\bibfield  {journal} {\bibinfo  {journal} {Optics communications}\ }\textbf
  {\bibinfo {volume} {114}},\ \bibinfo {pages} {353 } (\bibinfo {year}
  {1995})}\BibitemShut {NoStop}%
\bibitem [{\citenamefont {Carr}\ and\ \citenamefont {Leung}(2000)}]{Carr2000a}%
  \BibitemOpen
  \bibfield  {author} {\bibinfo {author} {\bibfnamefont {L.}~\bibnamefont
  {Carr}}\ and\ \bibinfo {author} {\bibfnamefont {M.}~\bibnamefont {Leung}},\
  }\href {http://stacks.iop.org/0953-4075/33/i=19/a=312} {\bibfield  {journal}
  {\bibinfo  {journal} {Journal of Physics B: Atomic, Molecular and Optical
  Physics}\ }\textbf {\bibinfo {volume} {33}},\ \bibinfo {pages} {3983}
  (\bibinfo {year} {2000})}\BibitemShut {NoStop}%
\bibitem [{\citenamefont {Malomed}(1998)}]{Malomed1998}%
  \BibitemOpen
  \bibfield  {author} {\bibinfo {author} {\bibfnamefont {B.}~\bibnamefont
  {Malomed}},\ }\href {http://pre.aps.org/abstract/PRE/v58/i2/p2564} {\bibfield
   {journal} {\bibinfo  {journal} {Physical Review E}\ }\textbf {\bibinfo
  {volume} {58}},\ \bibinfo {pages} {2564} (\bibinfo {year}
  {1998})}\BibitemShut {NoStop}%
\bibitem [{\citenamefont {Zhao}\ \emph {et~al.}(2017)\citenamefont {Zhao},
  \citenamefont {Alcala}, \citenamefont {McLain}, \citenamefont {Maeda},
  \citenamefont {Potnis}, \citenamefont {Ramos}, \citenamefont {Steinberg},\
  and\ \citenamefont {Carr}}]{PhysRevA.96.063601}%
  \BibitemOpen
  \bibfield  {author} {\bibinfo {author} {\bibfnamefont {X.}~\bibnamefont
  {Zhao}}, \bibinfo {author} {\bibfnamefont {D.~A.}\ \bibnamefont {Alcala}},
  \bibinfo {author} {\bibfnamefont {M.~A.}\ \bibnamefont {McLain}}, \bibinfo
  {author} {\bibfnamefont {K.}~\bibnamefont {Maeda}}, \bibinfo {author}
  {\bibfnamefont {S.}~\bibnamefont {Potnis}}, \bibinfo {author} {\bibfnamefont
  {R.}~\bibnamefont {Ramos}}, \bibinfo {author} {\bibfnamefont {A.~M.}\
  \bibnamefont {Steinberg}}, \ and\ \bibinfo {author} {\bibfnamefont {L.~D.}\
  \bibnamefont {Carr}},\ }\href {\doibase 10.1103/PhysRevA.96.063601}
  {\bibfield  {journal} {\bibinfo  {journal} {Phys. Rev. A}\ }\textbf {\bibinfo
  {volume} {96}},\ \bibinfo {pages} {063601} (\bibinfo {year}
  {2017})}\BibitemShut {NoStop}%
\bibitem [{\citenamefont {Potnis}\ \emph {et~al.}(2017)\citenamefont {Potnis},
  \citenamefont {Ramos}, \citenamefont {Maeda}, \citenamefont {Carr},\ and\
  \citenamefont {Steinberg}}]{PhysRevLett.118.060402}%
  \BibitemOpen
  \bibfield  {author} {\bibinfo {author} {\bibfnamefont {S.}~\bibnamefont
  {Potnis}}, \bibinfo {author} {\bibfnamefont {R.}~\bibnamefont {Ramos}},
  \bibinfo {author} {\bibfnamefont {K.}~\bibnamefont {Maeda}}, \bibinfo
  {author} {\bibfnamefont {L.~D.}\ \bibnamefont {Carr}}, \ and\ \bibinfo
  {author} {\bibfnamefont {A.~M.}\ \bibnamefont {Steinberg}},\ }\href {\doibase
  10.1103/PhysRevLett.118.060402} {\bibfield  {journal} {\bibinfo  {journal}
  {Phys. Rev. Lett.}\ }\textbf {\bibinfo {volume} {118}},\ \bibinfo {pages}
  {060402} (\bibinfo {year} {2017})}\BibitemShut {NoStop}%
\bibitem [{\citenamefont {Carr}\ \emph {et~al.}(2005)\citenamefont {Carr},
  \citenamefont {Holland},\ and\ \citenamefont
  {Malomed}}]{0953-4075-38-17-012}%
  \BibitemOpen
  \bibfield  {author} {\bibinfo {author} {\bibfnamefont {L.~D.}\ \bibnamefont
  {Carr}}, \bibinfo {author} {\bibfnamefont {M.~J.}\ \bibnamefont {Holland}}, \
  and\ \bibinfo {author} {\bibfnamefont {B.~A.}\ \bibnamefont {Malomed}},\
  }\href {http://stacks.iop.org/0953-4075/38/i=17/a=012} {\bibfield  {journal}
  {\bibinfo  {journal} {Journal of Physics B: Atomic, Molecular and Optical
  Physics}\ }\textbf {\bibinfo {volume} {38}},\ \bibinfo {pages} {3217}
  (\bibinfo {year} {2005})}\BibitemShut {NoStop}%
\end{thebibliography}%
\end{document}